\documentclass[12pt,a4paper]{article}
\pdfoutput=1
\usepackage{jheppub}
\usepackage{graphicx}
\usepackage{amssymb} 
\usepackage{amsmath,bm}
\usepackage{amsfonts}    
\DeclareMathOperator{\sech}{sech}
\DeclareMathOperator{\cosech}{cosech}

\newcommand\w[1]{\makebox[2.5em]{$#1$}}

\title{Holographic entanglement entropy for massive flavours in $dS_4$}

\author{Vladislav Vaganov} 
\affiliation{Department of Physics, Swansea University,\\Singleton Park, Swansea SA2 8PP, UK.}
\emailAdd{pyvv@swansea.ac.uk}
\abstract{We examine the holographic entanglement entropy of spherical regions in de Sitter space in the presence of massive flavour fields which are modelled by probe D7 branes in $AdS_5\times S^5$.      We focus on the finite part of the massive correction to the entropy in 
the limits of small mass and large mass that are separated by a phase transition between two topologically distinct brane embeddings.    For small masses, it approaches the flat space result for small spheres, whereas for large spheres there is a term that goes as the log of the sphere radius.   For large masses, we find evidence for a universal contribution logarithmic in the mass.   In all cases the entanglement entropy is smooth as the sphere radius crosses the horizon.}

\begin{document}
 \maketitle

\section{Introduction}
\label{sec:intro}
Entanglement entropy has emerged as a very useful tool for studying quantum field theories and many other areas of physics, see, for example, \cite{Calabrese:2004eu,Amico:2007ag,Kitaev:2005dm}.  

Whilst the physics of a conformal field theory in flat space is fairly featureless as a function of scale, interesting behaviour arises when new length or mass scales are added to the problem.    For instance, if we add a relevant deformation, or put the CFT in a curved background, or both, we expect to see interesting new effects such as quantum/thermal phase transitions \cite{Witten:1998qj}.  The entanglement entropy (EE) is a powerful tool for capturing this information \cite{Vidal:2002rm}.   A nice example showing both effects is a massive field theory in de Sitter space.   If we consider a region of size $R$ the EE may be expected to be sensitive to the two dimensionless combinations $R/l$ and $m l$, where $l$ is the de Sitter radius and $m$ the mass scale of the deformation.    

Computing EE in quantum field theories is difficult in general and has been done only for a certain number of simple cases, see, for example, \cite{Casini:2009sr}.    Fortunately, for field theories with a gravity dual, the computation of the EE at strong coupling is reduced to finding a minimal surface in the bulk that is homologous to the boundary of the entangling region, using the celebrated Ryu-Takayanagi (RT) formula \cite{Ryu:2006bv} or its covariant generalisation to extremal surfaces \cite{Hubeny:2007xt}. 

Whilst the holographic EE calculation is relatively straightforward in conformal backgrounds such as AdS, it becomes more challenging when we add a relevant deformation to the boundary CFT.    This corresponds to deforming the bulk geometry by some classical field and so not only do we first have to compute the backreaction, which can in itself be a highly nontrivial problem, we then have to find the minimal surfaces in the deformed geometry.    

Putting D-branes in $AdS_5\times S^5$ corresponds \cite{Karch:2002sh} in AdS/CFT to adding fundamental flavour fields to the dual $\mathcal{N}=4$ SYM and, when the flavour branes are separated from the D3 branes, the flavours are massive.   For quantities calculated directly from the on-shell action, it is not necessary to go beyond the probe limit $N_f \ll N_c$ in which backreaction is neglected.    This is not the case for flavour corrections to the EE, which are a topic of recent interest \cite{Jensen:2013lxa,Chang:2013mca,Kontoudi:2013rla,Karch:2014ufa,Jones:2015twa,Lewkowycz:2013laa,Bea:2013jxa,Georgiou:2015pia}.  However, \cite{Karch:2014ufa}, building on \cite{Lewkowycz:2013nqa}, found a shortcut to the problem which does not require computing the backreaction.    Their method is particularly convenient for spherical regions as one can exploit the map \cite{Casini:2011kv} from the EE to the thermal entropy on $\mathbb{R}\times H^{d-1}$, which is easily computed by the area of the horizon in a certain hyperbolic slicing of AdS.    Reference \cite{Karch:2014ufa} applied their method to compute the flavour correction to the EE for a spherical region in flat space, using the brane embedding of \cite{Karch:2002sh} in Poincar\'e $AdS_5\times S^5$.  

We apply the method of \cite{Karch:2014ufa} to study the EE of massive flavours in 4-dimensional de Sitter space of radius $l$, for a spherical region of proper radius $R$.   We focus in particular on the massive part of the flavour contribution to the EE.  There is also a CFT contribution from the massless flavours but it is not very interesting.    If $m$ is the flavour mass, the UV-finite part of the flavour contribution is a function of $R/l$ and $m l$.    We compute it in two different limits, one where the mass of the flavours is small compared to the de Sitter Hubble scale ($ml\ll 1$), and the opposite limit where it is large ($ml\gg 1$).   The calculation uses a D7 probe brane embedding in $dS_4$-sliced $AdS_5\times S^5$ which we have constructed perturbatively in both limits.   Similar embeddings preserving $\mathcal{N}=2$ supersymmetry were recently constructed analytically in \cite{Karch:2015vra,Karch:2015kfa}.   

The layout of this paper is as follows:  in Sec. \ref{sec:setup} we define the entangling region in de Sitter in two coordinate charts, and summarise what is already known about EE in 4d field theories in general and in de Sitter space.  In Sec. \ref{sec:cft} we look at the CFT case of empty AdS without branes, show how to map the calculation to the thermal entropy of a hyperbolic cylinder, and highlight the role played by the FRW region behind the horizon of the AdS bulk.   In Sec. \ref{sec:flavours} we construct the probe D7 brane embeddings both perturbatively for small and large masses and numerically for any mass, and locate the phase transition between the two topologically distinct embeddings.    In Sec. \ref{sec:flavourEE} we describe in detail the application of the method of \cite{Karch:2014ufa} to the calculation of the flavour EE and use the embeddings we found to compute the flavour EE in two different perturbative regimes.   In Sec. \ref{sec:conclude} we summarise our findings.   Appendix \ref{app:RT} describes the AdS extremal surfaces which compute the EE for a CFT in de Sitter and shows how this method agrees with the thermal entropy map.   Appendix \ref{app:metric} contains the metric determinant and its variation needed in Sec. \ref{sec:flavourEE}.

\section{Generalities}
\label{sec:setup}
\begin{figure}
\centering
\includegraphics[width= 2.5in]{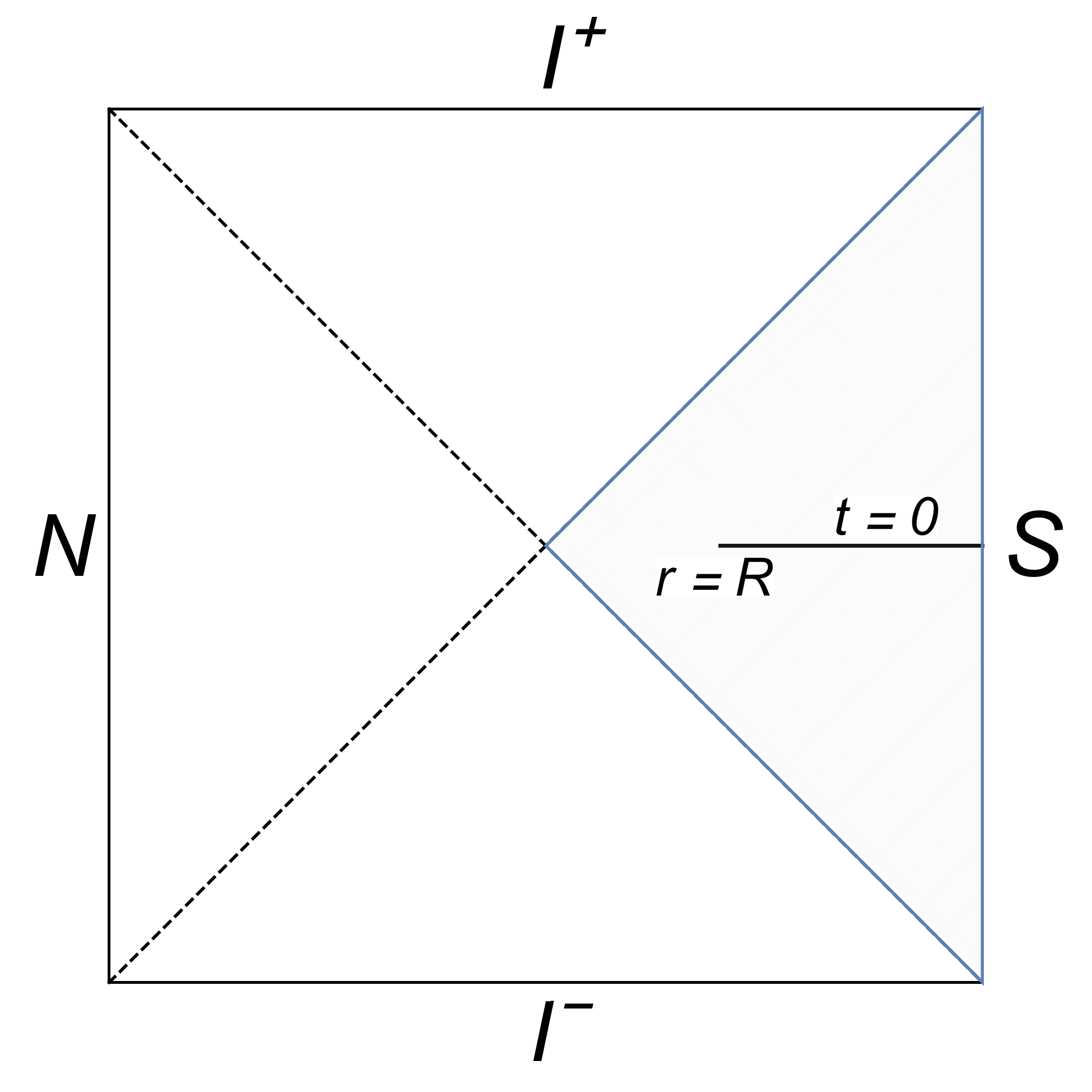}\,\,\includegraphics[width= 2.5in]{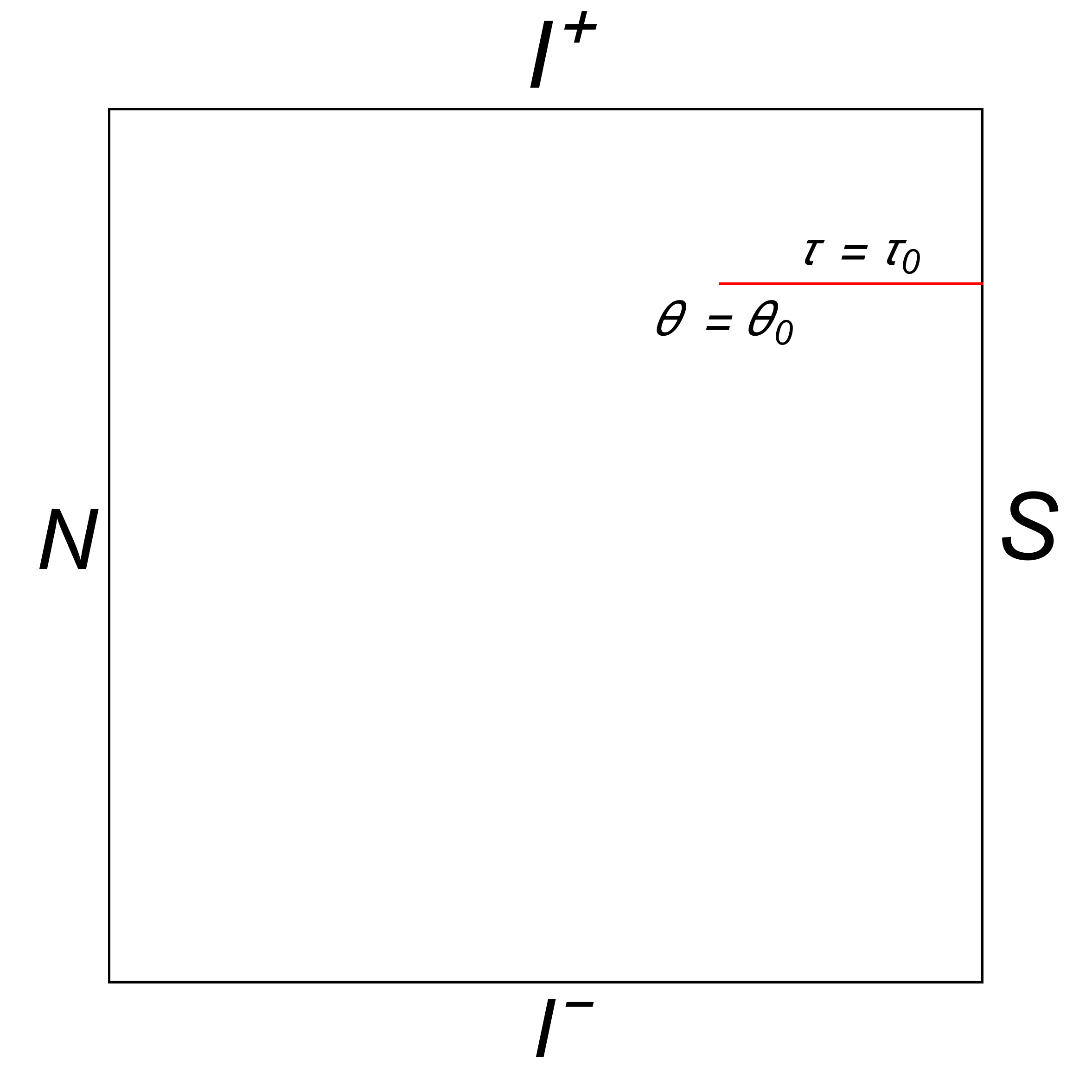}
\caption{Penrose diagrams of de Sitter space.  Left: the static patch (southern causal diamond) with an entangling region $0<r<R$ at time $t=0$.   Right: global de Sitter with an entangling region $0<\theta<\theta_0$ at time $\tau=\tau_0$.}
\label{fig:dsd}
\end{figure}

We are interested in computing the entanglement entropy (EE) for a spherical region in a quantum field theory living on $d$-dimensional de Sitter space of radius $l = H^{-1}$ where $H$ is the Hubble scale.     For an account of quantum field theory and coordinate charts in de Sitter, see \cite{Spradlin:2001pw}.    The main example we will work with is $dS_4$, however we keep the dimension general wherever possible.      The question can be phrased in a number of ways depending on the coordinate chart, see Fig. \ref{fig:dsd}, for example\footnote{We assume the field theory is in the standard Bunch-Davies vacuum state. }
\begin{enumerate}
\item
Suppose we have a field theory defined on the de Sitter static patch, which is the causal diamond accessible to an observer on the South pole,
with metric
\begin{equation}\label{eqn:setup1}
ds_{static}^2 = -\left(1-\frac{r^2}{l^2}\right)dt^2 + \frac{dr^2}{1-\frac{r^2}{l^2}}+r^2 d\Omega_{d-2}^2\,,
\end{equation}
where $t\in \mathbb{R}$ and $0\leq r\leq l$.   What is the EE of a spherical region $0<r<R$, where $R<l$, on time slice $t=0$?  
\item
Now suppose the field theory is defined on global de Sitter,
\begin{equation}\label{eqn:setup2}
ds_{global}^2 = -d\tau^2 + l^2 \cosh^2{(\tau/l)}\left(d\theta^2 +\sin^2{\theta}\,d\Omega_{d-2}^2\right),
\end{equation}
where $\tau\in\mathbb{R}$ and $0\leq \theta < \pi$. What is the EE of a spherical region of angular extent $0<\theta<\theta_0$ on time slice $\tau=\tau_0$?
\end{enumerate}
It is reasonable to expect that the answer to (2) should depend only on the proper radius of the entangling surface $R =  l \sin{\theta_0} \cosh{(\tau_0/l)}$.    When $R<l$ the entangling region is smaller than the horizon, whereas when $R>l$ it is bigger than the horizon.  
We also expect that the answer to (2) contains the answer to (1) when $R<l$, i.e. that in this domain the two EE's should both be given by the same function of $R/l$.   This is because the field theory on static de Sitter is obtained by tracing out degrees of freedom inaccessible to the static patch observer.    We will see how these expectations are justified in the holographic computations.

On general grounds we expect the EE of a spherical region of size $R$ in a 4d field theory to be the sum of a UV-divergent part and a UV-finite part.   In particular, for a CFT, we expect \cite{Ryu:2006ef}
\begin{equation}\label{eqn:gen1}
\mathcal{S} = a_1 \frac{R^2}{\epsilon^2} + a_2 \ln{(\epsilon/R)} + \mbox{other finite terms},
\end{equation}
where $\epsilon$ is the UV cut-off.      The UV divergent contributions are due to local effects, of which the first term is the ``area law'' \cite{Srednicki:1993im}.     The area law piece and the other finite terms depend on the regularisation scheme and the choice of vacuum state.  The coefficient $a_2$ of the logarithmic term is independent of the scheme and the state, and provides universal information that distinguishes different theories.   It is a conformally invariant combination of the central charges which appear in the trace anomaly, and geometrical factors involving the intrinsic and extrinsic curvatures of the entangling surface \cite{Solodukhin:2008dh}.    In particular, $a_2$ for a sphere in de Sitter is the same as $a_2$ for a sphere in flat space.     

More generally, when the CFT is deformed by a relevant operator, we expect, in addition to the terms in (\ref{eqn:gen1}), a universal logarithmic contribution \cite{Hertzberg:2010uv,Hung:2011ta,Lewkowycz:2012qr}
\begin{equation}\label{eqn:gen2}
\mathcal{S}_{univ} = a_3 \left(m R\right)^2 \ln{(m \epsilon)} ,
\end{equation}
where $m$ is the mass scale and the coefficient $a_3$ is independent of curvatures, and again provides universal information characterising the field theory.    In general, other terms with the coefficient of $\ln{(m \epsilon)}$ mixing curvatures and powers of the mass are expected but in 4d this is the only term \cite{Hung:2011ta}.   

Specialising now to 4d field theories in de Sitter, an important limit is when the entangling sphere is much bigger than the de Sitter horizon, $R/l \gg 1$.      In this case the overall finite part of the EE of a spherical region in $dS_4$ is expected to follow \cite{maldacenapimentel}
\begin{equation}\label{eqn:gen3}
\mathcal{S}_{finite} =  a_5 \frac{R^2}{l^2}+ a_6 \ln{(R/l)} + + \mbox{subleading},
\end{equation}
where the coefficient of the log term contains information about the long range entanglement of the state, related to particle creation effects in de Sitter. This term is not present in flat space.   For a CFT $a_6=-a_2$.   

Reference \cite{maldacenapimentel} further argued that when the mass scale of the relevant operator is small (compared to $H$), there is a non-zero $a_6$, whereas when the mass scale is large, or the theory has a mass gap, then $a_6 \simeq 0$, at least to leading order at large $N$.    For holographic theories on de Sitter the idea is that when the dual bulk geometry is ungapped, it contains another region that can be accessed by analytically continuing through a bulk horizon, whose metric is that of an expanding and contracting FRW cosmology.  Extremal surfaces measuring the EE of superhorizon sized regions in de Sitter probe this FRW part of the bulk, and in the limit $R\gg l$ the surfaces lies along the slice of maximal FRW scale factor, which gives a non-zero coefficient $a_6$ \cite{maldacenapimentel}.

When the entangling sphere is much smaller than the horizon, $R/l \ll 1$, we expect the EE in de Sitter to be the same as in flat space.   

We will see that for the examples we study in this paper, the CFT in Sec. \ref{sec:cft} and the relevant deformation corresponding to the addition of massive flavours in Sec. \ref{sec:flavourEE}, the EE agrees with the above expectations.

\section{CFT entanglement entropy from CHM map} 
\label{sec:cft}
We will begin by addressing the above questions for the CFT without flavours in de Sitter space.    We show how to exploit the Casini, Huerta, Myers (CHM) method \cite{Casini:2011kv} to map the problem to a simple horizon area calculation.   Although the CFT EE can also be computed by a direct application of the Ryu-Takayanagi formula (refer to Appendix \ref{app:RT}), the flavour EE calculation we will do in Sec. \ref{sec:flavourEE} relies on the CHM method, so it is important to understand how it works for the CFT.   

CHM demonstrated that for a CFT in Minkowski space $\mathbb{R}^{1,d-1}$ or cylindrical space $\mathbb{R}\times S^{d-1}$ one can map the problem of computing the EE of a spherical region to the problem of computing the thermal entropy on a geometry which is the direct product of time with hyperbolic space $H^{d-1}$.   The thermal entropy is just the area of the horizon in a hyperbolic slicing of AdS.    We show how to apply the CHM method to the case where the boundary is de Sitter space.   

The result (\ref{eqn:eecft4}) for the EE of a CFT in $dS_4$ has appeared in the literature before \cite{Fischler:2013fba}, derived from an extremal surface.     The new elements in this section are the derivation of this result from the CHM map in two coordinate charts (static patch and global de Sitter) and the conclusion that (in global de Sitter, where it makes sense to talk of superhorizon sized spheres),  (\ref{eqn:eecft4}) continues to be valid when the sphere is superhorizon sized i.e. $R>l$.    

We emphasise that the EE of superhorizon sized regions probes the FRW geometry behind the horizon of the de Sitter slicing of AdS, as noted in \cite{maldacenapimentel}.    Reference \cite{Fischler:2013fba} argued that the EE of superhorizon sized regions is given by an expression different from (\ref{eqn:eecft4}) and concluded that the EE goes through a phase transition when $R$ crosses the horizon.    As noted in \cite{Fischler:2013fba}, the reason for the discrepancy is that for a superhorizon sized entangling sphere, \cite{Fischler:2013fba} use a non-smooth (or discontinuous) Ryu-Takayanagi surface that does not enter the FRW region.\footnote{Their surface can be viewed either as a disconnected surface or as a connected surface consisting of two disconnected pieces extending as far as the horizon $z=2 l$ plus a part of the horizon.  We thank Juan Pedraza for clarifying this.}     We make some comments about this in Sec. \ref{subsec:horizon}.

\subsection{Hyperbolic slicing of AdS} 
We show how to apply the CHM method to the case where the boundary is de Sitter space.   We begin by introducing the hyperbolic slicing of Lorentzian anti de Sitter space $AdS_{d+1}$ of radius $L$, which can be described by the quadric
\begin{equation}
-y_{-1}^2 - y_0^2 +y_1^2 +...+y_d^2 = -L^2
\end{equation}
embedded in $\mathbb{R}^{2,d}$ with metric
\begin{equation}
ds^2 = -dy_{-1}^2 -dy_0^2 +dy_1^2+...+dy_d^2.
\end{equation}
The hyperbolic slicing of AdS is given by
\begin{align}\label{eqn:hyperbolic slicing}
& y_{-1}= \rho \cosh{u}\qquad y_0 =\sqrt{\rho^2-L^2} \sinh{(\sigma/L)} \nonumber \\
& y_i =\rho\sinh{u}\, n_i\ \qquad y_d = \sqrt{\rho^2-L^2} \cosh{(\sigma/L)} 
\end{align} 
where $\rho\geq L$, $u\geq 0$, $\sigma\in \mathbb{R}$ and the $n_i$ here and henceforth denotes the components of a $(d-1)$-dimensional unit vector.     The metric is
\begin{equation}\label{eqn:hyperbolic metric}
ds^2 = -\left(\frac{\rho^2}{L^2}-1\right)d\sigma^2+ \frac{d\rho^2}{\frac{\rho^2}{L^2}-1} +\rho^2 \left(du^2 + \sinh^2{u}\, d\Omega_{d-2}^2\right),
\end{equation} 
which can be interpreted as a topological black hole with a horizon at $\rho=L$ with $H^{d-1}$ spatial geometry.    Taking the asymptotic limit $\rho\to\infty$ and removing a factor of $\rho^2/L^2$ the CFT lives on the boundary metric
\begin{equation}\label{eqn:open}
ds_b^2 = -d\sigma^2 + L^2 \left(du^2 + \sinh^2{u}\, d\Omega_{d-2}^2\right)
\end{equation}
which is the open Einstein universe (or hyperbolic cylinder) $\mathbb{R}\times H^{d-1}$.     

The thermal entropy of the boundary CFT is proportional to the area of the horizon in the bulk:
\begin{equation}\label{eqn:thermal entropy}
\mathcal{S}_{thermal}  =\frac{L^{d-1}V_{S^{d-2}}}{4 G_N} \int_0^{u_{\infty}} \sinh^{d-2}{u}\, du,
\end{equation} 
where $G_N$ is the $d+1$-dimensional Newton constant and we have introduced an IR regulator $u_{\infty}$ that we will define carefully later.  

Isometries of $AdS_{d+1}$ correspond to rotations and boosts in the embedding space.    These can be used to generate equivalent foliations that we will find useful.  In particular, a boost through parameter $\beta$ in the $(y_{-1},y_d)$ plane followed by a rotation through angle $\alpha$ in the $(y_{-1},y_0)$ plane is given by
\begin{equation}\label{eqn:boosted hyperbolic slicing}
\left( \begin{array}{c} y_{-1} \\ y_0 \\ y_i \\ y_d \end{array} \right) \rightarrow \begin{pmatrix} \cos{\alpha} & -\sin{\alpha} & 0 & 0 \\ \sin{\alpha} & \cos{\alpha} & 0 & 0 \\ \w0 & \w0 & \w1 & \w0 \\ 0 & 0 & 0 & 1 \end{pmatrix} \begin{pmatrix} \cosh{\beta} & 0 & 0 & \sinh{\beta} \\ \w0 & \w1 & \w0 & \w0 \\ 0 & 0 & 1 & 0 \\ \sinh{\beta} & 0 & 0 & \cosh{\beta} \end{pmatrix}  \left( \begin{array}{c} y_{-1} \\ y_0 \\ y_i \\ y_d \end{array} \right).
\end{equation}
where the $y_{\mu}$ are (\ref{eqn:hyperbolic slicing}).   The metric (\ref{eqn:hyperbolic metric}) is unchanged.    However, as we will see below, it affects the way the bifurcation surface $\rho=L$ intersects the boundary from the perspective of other coordinate charts, and hence it affects the regulator $u_\infty$ in (\ref{eqn:thermal entropy}) and the entropy.   

\subsection{EE in static de Sitter} 
\label{subsec:staticdesitter}
We are going to take the generalized hyperbolic slicing (\ref{eqn:boosted hyperbolic slicing}) with a boost $\beta$ and no rotation, and show that the whole $\mathbb{R}\times H^{d-1}$ boundary is mapped to the causal development of the spherical region $0<r<R=l \sech{\beta}$ in static de Sitter slicing.   From this it will follow that we can compute the EE of this spherical region by evaluating the thermal entropy of $\mathbb{R}\times H^{d-1}$ which is given by (\ref{eqn:thermal entropy}).

In Fefferman-Graham form the static de Sitter slicing of AdS is 
\begin{align}\label{eqn:static slicing}
y_{-1} &= \frac{L\, l}{z}\left(1+\frac{z^2}{4l^2}\right)\quad &&y_0  = \frac{L\, l}{z}\left(1-\frac{z^2}{4l^2}\right) \sqrt{1-r^2/l^2} \sinh{(t/l)} \nonumber \\
y_i &=  \frac{L \,l}{z}\left(1-\frac{z^2}{4l^2}\right) \frac{r}{l}\, n_i\quad &&y_d =  \frac{L \,l}{z}\left(1-\frac{z^2}{4l^2}\right)  \sqrt{1-r^2/l^2}\cosh{(t/l)} ,
\end{align}
where $0\leq r \leq  l$ and $t\in \mathbb{R}$, with line element 
\begin{equation}\label{eqn:static slicing metric}
ds^2 = \frac{L^2}{z^2}dz^2 + \frac{L^2}{z^2}f(z) \left(-\left(1-\frac{r^2}{l^2}\right)dt^2 + \frac{dr^2}{1-\frac{r^2}{l^2}}+r^2 d\Omega_{d-2}^2\right),
\end{equation}
where $f(z) = (1-z^2/4l^2)^2$.      The asymptotic boundary is at $z=0$ and there is a horizon in the bulk at $z= 2l$.    Taking the limit $z\to 0$ and removing a factor of $L^2/z^2$ the CFT lives on the static patch of de Sitter (\ref{eqn:setup1}) of radius $l$.    

To determine the entangling surface on the boundary we find where the bifurcation surface $\rho=L$ of the hyperbolic foliation (with a boost $\beta$) intersects the boundary in terms of the static de Sitter coordinates.  Observe that
\begin{equation}\label{eqn:observe}
\frac{L z}{2l} =  y_{-1}-\sqrt{y_{-1}^2-L^2} ,\qquad  \tanh{(t/l)} = \frac{y_0}{y_d},\qquad \frac{l^2}{r^2}-1= \frac{y_d^2-y_0^2}{|y_i|^2},
\end{equation}
where $|y_i|^2 \equiv \sum_{i=1}^{d-1} y_i^2$.    The embedding coordinates of the bifurcation surface $\rho=L$ in the boosted hyperbolic foliation are
\begin{align}
y_{-1} &= L\cosh{\beta}\cosh{u}\qquad &&y_0 = 0 \nonumber \\
y_i  &= L\sinh{u}\,n_i\qquad &&y_d = L\sinh{\beta}\cosh{u}.
\end{align}
Taking the limit $u\to\infty$ we find the intersection of this bifurcation surface with the boundary in terms of the static de Sitter coordinates:
\begin{equation}
\tanh{(t/l)} = 0,\qquad \frac{l^2}{r^2}-1 = \sinh^2{\beta}.
\end{equation}
It follows that the entangling surface in de Sitter is a sphere of radius $R=l\sech{\beta}$ at $t=0$, as we set out to show.   This suggests that 
the map from the EE of this spherical region to the thermal entropy on $\mathbb{R}\times H^{d-1}$ proceeds in the same way as in the examples of flat space and cylindrical space $\mathbb{R}\times S^{d-1}$ in \cite{Casini:2011kv}, but to be rigorous we need to check a couple of other conditions.   From (\ref{eqn:observe}) get the restriction of the bulk transformation $(\sigma,\rho,u)\to (z,t,r)$ to the boundary:
\begin{align}\label{eqn:btransdesitter}
\tanh{(t/l)} &= \frac{\sinh{(\sigma/L)}}{\cosh{\beta}\cosh{(\sigma/L)} +\sinh{\beta}\cosh{u}},  \nonumber \\
\frac{r}{l}& = \pm\frac{\sinh{u}}{\cosh{\beta}\cosh{u}+\sinh{\beta}\cosh{(\sigma/L)}},
\end{align} 
It maps the static de Sitter metric (\ref{eqn:setup1}) to 
\begin{align}\label{eqn:modulo}
&ds^2  = \Omega^2 \left[-\frac{l^2}{L^2}d\sigma^2 + l^2\left(du^2+\sinh^2{u}\,d\Omega_{d-2}^2\right)\right]\nonumber \\
& \Omega = \left(\cosh{\beta}\cosh{u}+\sinh{\beta}\cosh{(\sigma/L)}\right)^{-1},
\end{align}
which is the metric of an $\mathbb{R}\times H^{d-1}$ of size $l$, (c.f. (\ref{eqn:open})) times a conformal factor.   We can further show that the coordinates $(\sigma,u)$ cover precisely the causal development $\mathcal{D}$ of the ball inside the entangling surface $r=R$ at $t=0$ in static de Sitter.\footnote{One way to see this is to note that the future causal development is bounded by the null ray
\begin{equation}
\tanh{(t/l)} = \frac{R/l-r/l}{1-R r/l^2},
\end{equation}
and graphically plot the set of points generated by the boundary transformation (\ref{eqn:btransdesitter}) to see that it is bounded by this null ray.}     We would also need to show that the vacuum correlators on $\mathcal{D}$ transform under the conformal mapping to thermal correlators on $\mathbb{R}\times H^{d-1}$ with temperature $1/2\pi l$, and hence that there is a unitary map between the reduced density matrix on $\mathcal{D}$ and the thermal density matrix on $\mathbb{R}\times H^{d-1}$ from which the result would follow by virtue of the invariance of the von Neumann entropy under unitary transformations.     We have not checked this carefully but we assume it works in the same way as in \cite{Casini:2011kv}, at least for the Bunch-Davies vacuum in de Sitter.    The result we find for the CFT EE is the same as we get by a direct application of the Ryu-Takayanagi formula in the de Sitter slicing coordinates in Appendix \ref{app:RT}.   

It remains to determine the IR regulator $u_\infty$ to use in the thermal entropy integral (\ref{eqn:thermal entropy}).   We need to match it to the short distance cut-off $\epsilon$ in the boundary CFT.    The standard holographic relationship between $\epsilon$ and the minimum value of the Fefferman-Graham radial coordinate is $z_{min}=\epsilon$.    On the horizon of the $\beta$-boosted hyperbolic foliation 
\begin{equation}
y_{-1} = L \cosh{\beta} \cosh{u} = \frac{L l}{z}\left(1+\frac{z^2}{4l^2}\right).
\end{equation}
Using $R=l\sech{\beta}$ we find the regulator to use for the horizon area calculation is
\begin{equation}\label{eqn:dscutoff}
\cosh{u_{\infty}} = \frac{R}{\epsilon}\left(1+\frac{\epsilon^2}{4 l^2}\right).
\end{equation}
For the CFT in  $dS_4$ this gives an EE of\footnote{Note that in the holographic correspondence between $AdS_5\times S^5$ and $\mathcal{N}=4$ SYM,  $L^3/G_N = 2 N^2/\pi$. }
\begin{equation}\label{eqn:eecft4}
\mathcal{S}_{EE}^{CFT_4} = \frac{L^3 V_{S^2}}{4 G_N} \int_0^{u_{\infty}} \sinh^2{u}\,du  = \frac{L^3 V_{S^2}}{16G_N} \left(\frac{2R^2}{\epsilon^2} - 2\ln{(2R/\epsilon)} + \frac{ R^2}{l^2}-1\right).
\end{equation} 
The area law divergence and the log term are as in flat space and consistent with expectations (\ref{eqn:gen1}).   The piece $R^2/l^2$ does not appear in flat space; its coefficient is scheme-dependent.   This expression agrees with \cite{Fischler:2013fba}.   

\subsection{EE in global de Sitter} \label{sec:globaldesitterfoliation}
We can repeat the same trick for global de Sitter.    As we want to vary both the size $\theta_0$ of the entangling surface and the time slice $\tau_0$, we will use the generalized hyperbolic slicing (\ref{eqn:boosted hyperbolic slicing}) with boost $\beta$ and rotation $\alpha$.   We will show that the entire $\mathbb{R}\times H^{d-1}$ boundary is mapped to the causal development of the spherical region $0<\theta<\theta_0=\tan^{-1}{(\cosech{\beta})}$ at time $\tau_0 = l \tanh^{-1}{(\sin{\alpha})}$ in global de Sitter.   As expected, the EE depends only on the combination $R =  l \sin{\theta_0} \cosh{(\tau_0/l)}$ which is the proper radius of the entangling surface in global de Sitter.  

The global de Sitter slicing of AdS in Fefferman-Graham form is 
\begin{align}\label{eqn:global slicing}
y_{-1} &= \frac{L l}{z}\left(1+\frac{z^2}{4l^2}\right) \quad &&y_0 =\frac{L l}{z}\left(1-\frac{z^2}{4 l^2}\right) \sinh{(\tau/l)}\nonumber \\
y_i &= \frac{L l}{z}\left(1-\frac{z^2}{4l^2}\right) \cosh{(\tau/l)} \,n_i \sin{\theta} \quad &&y_d = \frac{L l}{z}\left(1-\frac{z^2}{4l^2}\right) \cosh{(\tau/l)} \cos{\theta} ,
\end{align}
where $0\leq \theta <\pi$ and $\tau\in \mathbb{R}$, with line element
\begin{equation}\label{eqn:global slicing metric}
ds^2 =  \frac{L^2}{z^2}dz^2+\frac{L^2}{z^2}f(z)\left(-d\tau^2 + l^2 \cosh^2{(\tau/l)} (d\theta^2+\sin^2{\theta}\,d\Omega_{d-2}^2)\right) ,
\end{equation}
where as in Sec. \ref{subsec:staticdesitter}  we have $f(z) = (1-z^2/4l^2)^2$, the asymptotic boundary is at $z=0$ and the horizon in the bulk is at $z= 2l$.   The CFT lives on global de Sitter space of radius $l$.  The global de Sitter slicing coordinates can be expressed in terms of the embedding coordinates as
\begin{equation}\label{eqn:observe global}
\frac{L z}{2l} =  y_{-1}-\sqrt{y_{-1}^2-L^2},\qquad \tanh{(\tau/l)} = \frac{y_0}{\sqrt{|y_i|^2+y_d^2}},\qquad \tan{\theta} = \frac{|y_i|}{y_d}.
\end{equation}
The embedding coordinates of the bifurcation surface $\rho=L$ in the boosted and rotated hyperbolic foliation are
\begin{align}\label{eqn:bifurcation}
y_{-1} &= L\cos{\alpha}\cosh{\beta}\cosh{u}\qquad &&y_0 = L\sin{\alpha}\cosh{\beta}\cosh{u} \nonumber \\
y_i  &= L\sinh{u}\,n_i\qquad &&y_d = L\sinh{\beta}\cosh{u}.
\end{align}
Taking the boundary limit $u\to\infty$ it follows that in the global de Sitter chart the bifurcation surface intersects the boundary at 
\begin{equation}\label{eqn:globaldsintersection}
\tanh{(\tau_0/l)} = \sin{\alpha},\qquad \tan{\theta_0} = \cosech{\beta}.
\end{equation} 
It follows that the entangling surface in de Sitter is a sphere of angular size $\theta_0=\tan^{-1}{(\cosech{\beta})}$ at time $\tau_0 = l \tanh^{-1}{(\sin{\alpha})}$.   With some more work we can show, as we did in Sec. \ref{subsec:staticdesitter}, by looking at the transformation between boundary coordinates, that the entire $\mathbb{R}\times H^{d-1}$ boundary is mapped to the causal development of the region inside the entangling surface.  It follows (modulo caveats below (\ref{eqn:modulo})) that the EE is equal to the thermal entropy on $\mathbb{R}\times H^{d-1}$ evaluated with the appropriate regulator.    Using the same procedure as in the static de Sitter case we can relate the CFT short distance cut-off $\epsilon$ to the regulator $u_{\infty}$.   It takes exactly the same form as (\ref{eqn:dscutoff}) with the replacement
\begin{equation}\label{eqn:above}
R =  l \sin{\theta_0} \cosh{(\tau_0/l)}.   
\end{equation}
For the CFT in $dS_4$ this again gives the expression (\ref{eqn:eecft4}) for the holographic EE where $R$ is now given by (\ref{eqn:above}).    We see that it does not depend on $\theta_0$ and $\tau_0$ independently, but only through the combination $R$.   It agrees with the expectation in Sec. \ref{sec:setup} that the EE in global de Sitter for subhorizon sized regions agrees with the EE in static de Sitter.\footnote{This is similar to the observation in \cite{Fischler:2013fba} that the EE of a sphere in the static patch is the same as the EE of a sphere in the conformally flat patch of de Sitter of the same proper radius, provided the radius is smaller than the size of the horizon.}   Further, this derivation indicates that the EE is given by (\ref{eqn:eecft4}) for all $R/l$ and is smooth as $R$ crosses the horizon.     To confirm this, we have computed the extremal surface in the global de Sitter slicing coordinates and applied the Ryu-Takayanagi formula in Appendix \ref{app:RT}. It gives the same result, modulo a subtlety that we discuss below.   

A simple example of a superhorizon sized entangling surface is an equatorial surface at late times.    Putting $\theta_0 = \pi/2$ and taking $\tau_0\to\infty$ we get 
\begin{equation}\label{eqn:latetimelimit}
\mathcal{S}_{EE}^{CFT_4}  = \frac{L^3 V_{S^2}}{16 G_N}\left(\left(\frac{l^2}{2\epsilon^2} +\frac{1}{4}\right)e^{2\tau_0/l} +\left(\frac{l^2}{\epsilon^2}-\frac{1}{2}\right)- 2\ln{(l/\epsilon)} - \frac{2 \tau_0}{l}\right) + \mathcal{O}(e^{-2\tau_0/l}).
\end{equation}
This agrees with \cite{maldacenapimentel} who derived it from a late time RT surface, as it must, since the RT method agrees with the CHM method for arbitrary size spherical entangling regions, as we show in Appendix \ref{app:RT}.  

Ref. \cite{Fischler:2013fba} also computed the holographic RT surfaces that measure the EE of a spherical region in a CFT in de Sitter space in various dimensions.   They argued that there is a phase transition when the size $R$ of the sphere crosses the horizon.   They noted that the reason for the discrepancy between their result and the late time limit of \cite{maldacenapimentel} is that the surface they used for superhorizon sized regions was non-smooth (or disconnected), whilst the surface of \cite{maldacenapimentel} is smooth.\footnote{This non-smooth/disconnected surface has a smaller area than the smooth surface.}   The extremal surfaces we use (see Appendix \ref{app:RT}) agree with \cite{maldacenapimentel}, are smooth and connected, and show no evidence for a phase transition.    This highlights an important point which we now address.

\subsection{Behind the horizon in the de Sitter slicing} 
\label{subsec:horizon}
The $dS_d$ slicing of Lorentzian $AdS_{d+1}$ (\ref{eqn:global slicing metric}) can be obtained from the $S^d$ slicing of Euclidean $AdS_{d+1}$ 
\begin{equation}
ds^2 = \frac{L^2}{z^2}dz^2 + \frac{L^2}{z^2}f(z)\, l^2 d\Omega_4^2 
\end{equation} 
by a Wick rotation of the polar angle on the $S^d$ slice.\footnote{The static $dS_d$ slicing (\ref{eqn:static slicing metric}) can be obtained from a Wick rotation of another coordinate on $S^d$ e.g. for $S^2$ it is the azimuthal coordinate, for $S^3$ it is one of the angles in the Hopf coordinates.    This slicing also has a continuation behind the bulk horizon.}     It covers only a part of global $AdS_{d+1}$ as indicated by the presence of a horizon at $z= 2l$ which in the Euclidean slicing was the origin.     To continue through this horizon into another patch, we first switch to another radial coordinate by writing $z= 2l e^{-\xi/L}$.  The Lorentzian metric (\ref{eqn:global slicing metric}) takes the form
\begin{equation}
ds^2 = d\xi^2 + \sinh^2{(\xi/L)}\left(\frac{L}{l}\right)^2 \left(-d\tau^2 + l^2 \cosh^2{(\tau/l)} \,d\Omega_{d-1}^2\right) 
\end{equation}
The spacetime behind the horizon can now be accessed by analytically continuing the coordinates $\xi$ and $\tau$:
\begin{equation}
\xi \to i\hat{\xi},\qquad \tau  = -\frac{i\pi l}{2} + \hat{\tau}
\end{equation}
In this region the metric describes an FRW geometry with hyperbolic ($H^d$) spatial slices: 
\begin{equation}
ds^2 = -d\hat{\xi}^2 + \sin^2{(\hat{\xi}/L)}\left(\frac{L}{l}\right)^2 \left(d\hat{\tau}^2 + l^2 \sinh^2{(\hat{\tau}/l)} \,d\Omega_{d-1}^2\right) 
\end{equation}
The scale factor increases from zero at the horizon, attains a maximum at $\hat{\xi}/L=\pi/2$ and vanishes again at $\hat{\xi}/L=\pi$.   

The EE of superhorizon sized regions in de Sitter probes this FRW geometry \cite{maldacenapimentel}.     To see this in the CHM method, observe that the integral (\ref{eqn:thermal entropy}) used to compute the EE has a lower limit $u=0$.     Using (\ref{eqn:observe global}) and (\ref{eqn:bifurcation}) we find that along the bifurcation surface the integral is defined over, each point $u$ corresponds to a point
\begin{equation}
z = 2l^2 R^{-1}\left(\cosh{u}-\sqrt{\cosh^2{u}-R^2/l^2}\right)
\end{equation}
in the $dS_d$ slicing.    In particular, if $R>l$, the lower limit $u=0$ corresponds to a complex $z$, which can be seen to map to a real $\hat{\xi}>0$.    If we send $R/l\to\infty$, the lower limit maps to the maximal scale factor slice $\hat{\xi}/L=\pi/2$.    

Of course the thermal entropy integral in the CHM method is precisely the integral computing the area of the extremal surface in the $dS_d$ slicing written in a more convenient coordinate system.    So the lower limit $u=0$ corresponds to the maximal extent of the extremal surface along the radial $z$ direction.  

\paragraph{Note added in version 2} 
Whilst the non-smooth/disconnected RT surfaces in \cite{Fischler:2013fba} may have some role to play, several arguments suggest that the smooth surfaces described in Appendix \ref{app:RT} of this paper (which agree with the CHM map method and with \cite{maldacenapimentel}), correctly give the EE of superhorizon sized spheres in the CFT, and thus that there is no phase transition in the EE as the sphere size crosses the horizon:  
\begin{itemize}
\item
Conformal invariance of the logarithmic term in (\ref{eqn:gen1}).   If we use the non-smooth surface this term is missing, i.e. there is no long-range entanglement term $\ln{(R/l)}$.  
\item
The absence of the $\ln{(R/l)}$ piece would also be surprising given that it appears in the theory of a free massive scalar field, as shown in \cite{maldacenapimentel}.  
\item
There is a unique and well-defined answer for the RT surface corresponding to a spherical entangling region on the $\mathbb{R}\times S^{d-1}$ boundary of global $AdS_{d+1}$, given in Appendix \ref{app:RT}.   When transformed to the de Sitter slicing of AdS, it gives the smooth surface.   
\item
It is known that holographic field theories in de Sitter have the property that two-point correlators, in the geodesic limit, for superhorizon sized separations, are sensitive to the FRW geometry behind the bulk horizon, see e.g., \cite{Kumar:2015jxa}, and also see e.g.  \cite{Festuccia:2005pi}, for the analogous situation involving thermal field theories dual to AdS black holes.   It would be surprising if there was such a marked difference between the two geometric probes (i.e. EE and correlators in the geodesic limit).   
\end{itemize}
We note that the result of \cite{Fischler:2013fba} for the EE of superhorizon sized spheres of a CFT in $dS_4$ can be reproduced in our formalism by cutting off the integral ((\ref{eqn:thermal entropy}) or (\ref{eqn:smin})) at the bulk horizon $z=2 l$.

\section{Flavours in de Sitter} 
\label{sec:flavours}  
To add massive flavour fields to $dS_4$ we put probe D7 branes in $dS_4$-sliced $AdS_5\times S^5$.   We will do this by working in Euclidean signature and solving for smooth embeddings in $S^4$-sliced Euclidean $AdS_5\times S^5$.   As we will see, the equations of the embedding depend only on the $S^3$ slipping mode and the warp factor multiplying the $S^4$, and are insensitive to the metric on the slice.   Thus the solutions can be continued to the de Sitter slicing.\footnote{After submitting this paper we learned that these embeddings were studied in \cite{Hirayama:2006jn}.}    We will use the solutions so obtained in Sec. \ref{sec:flavourEE} to compute the entanglement entropy contribution of the flavours. 

\subsection{Probe D7 branes} 
We take the Euclidean $AdS_5\times S^5$ metric in the Fefferman-Graham form
\begin{align}
g & =g_{AdS_5} + g_{S^5} \nonumber \\ 
&= \frac{L^2}{z^2}dz^2 + \frac{L^2}{z^2}f(z)\, l^2 d\Omega_4^2 + L^2\left(d\psi^2 + \cos^2{\psi}\,d\theta^2+\sin^2{\psi}\,d\Omega_3^2\right),
\end{align}
where $f(z) = (1-z^2/4l^2)^2$ and we are thinking of the $S^4$ slices as having radius $l$.    We consider probe $D7$ branes wrapping the whole of $AdS_5$ and the $S^3$ inside the $S^5$, sitting at $\theta=0$ and having $\psi(z)$ as the slipping mode.  The induced metric on the brane is
\begin{equation}\label{eqn:induced metric}
\gamma = L^2\left(\frac{1}{z^2}+\psi'^2(z)\right)dz^2 +  \frac{L^2}{z^2}f(z) \,l^2 d\Omega_4^2 + L^2\sin^2{\psi(z)}\,d\Omega_3^2.
\end{equation}
The D7 brane geometry can be viewed a cone with an $S^4\times S^3$ base, with the $S^4$ the radial slice in AdS and the $S^3$ a part of the internal space.      The spheres shrink as we move inward along the radial direction, and we can have two topologically distinct classes of solution depending on which sphere caps off first, as we will explain.     

The brane action is the Dirac-Born-Infeld (DBI) action 
\begin{equation}\label{eqn:dbi}
S_{brane}= -T_7 \int \sqrt{\mbox{det}\,\gamma} \, d^8 x = -T_7 (L l)^4 V_{S^4} V_{S^3} \int dz \left(\frac{L}{z}\right)^4 f(z)^2 \sin^3{\psi}\, \sqrt{\frac{1}{z^2}+\psi'^2}
\end{equation}  
where $T_7$ is the brane tension.   It is just the volume of the brane.    We do not turn on a worldvolume gauge field so we do not expect the embeddings to preserve any supersymmetries.   Supersymmetric embeddings of this type with a gauge field were recently constructed analytically in \cite{Karch:2015vra,Karch:2015kfa}.  Extrema of (\ref{eqn:dbi}) satisfy the second-order equation
\begin{align}\label{eqn:embedding ode}
& z\sin^3{\psi}\,\left(z(z^2-4l^2)\psi''+ 4z^2(4l^2+z^2)\psi'^3+(12l^2+5 z^2)\psi'\right)+ \nonumber \\
&\qquad\qquad\qquad\qquad\qquad+3\sin^2{\psi}\cos{\psi}\,(4l^2-z^2)(1+z^2\psi'^2)=0.
\end{align}
We will shortly solve this equation numerically and in a perturbation series about two different limits, for solutions that we require to be smooth in the IR.    First we discuss holographic renormalisation, quoting results from \cite{Karch:2005ms}.\footnote{In the discussion of holographic renormalisation we use units $L=l=1$ for clarity.}   Linearising the equation about $\psi=\pi/2$ gives a scalar dual to an operator of dimension $3$.  The source for this operator, identified with the flavour mass, and the VEV, identified with the chiral condensate, can be extracted from the asymptotic expansion of the scalar $\phi = \frac{\pi}{2}-\psi$
\begin{equation}
\phi  = \phi_0 z + \phi_2 z^3 +\frac{1}{12}R_0 \phi_0 z^3 \ln{z} +...,
\end{equation}
where $\phi = \frac{\pi}{2}-\psi$ and $R_0=12$ is the Ricci scalar of the leading term $h_0$ in the near-boundary expansion of the AdS part of the metric
\begin{align}
& g_{AdS_5} = \frac{1}{z^2}dz^2 + \frac{1}{z^2} h \nonumber \\
&h = f(z) h_0,\qquad h_0 = d\Omega_4^2.
\end{align}
If we denote by $g_{\epsilon} = \epsilon^{-2} f(\epsilon) h_0$ the metric on the regulator surface $z=\epsilon$, the counterterms for this system are \cite{Karch:2005ms}
\begin{align}\label{eqn:counterterms}
L_1 &= -\frac{1}{4} \sqrt{\mbox{det}\,g_\epsilon}\left(1-\frac{R_\epsilon}{12}+\frac{\ln{\epsilon}}{8}\left(R^\epsilon_{ij}R_\epsilon^{ij}-\frac{1}{3}R_\epsilon^2\right)\right) \nonumber \\
L_2 &= \frac{1}{2}\sqrt{\mbox{det}\,g_\epsilon}\left(\phi^2_\epsilon + \phi_\epsilon\Box^\epsilon_W \phi_\epsilon \ln{\phi_\epsilon}\right) \nonumber \\
L_f & = -\frac{5}{12}\sqrt{\mbox{det}\,g_\epsilon} \,\phi^4_\epsilon .
\end{align}
where $\Box_W^\epsilon = \Box^\epsilon + R_\epsilon/6$ is the Weyl-covariant Laplacian.      The choice of the modified scalar field counterterm ($\ln{\phi_\epsilon}$ instead of the usual $\ln{\epsilon}$) and scheme for the finite counterterms are motivated by the requirement that in the large mass limit we should recover the flat space result, in particular, the VEV should vanish.     The logic is the same as for the similar embeddings in global AdS constructed in \cite{Karch:2006bv}.   If $S_{brane,\epsilon}$ is the DBI action (\ref{eqn:dbi}) evaluated up to $z=\epsilon$, the subtracted action is
\begin{align}
S_{sub} &= S_{brane,\epsilon}+S_{ct,\epsilon}, \nonumber \\
S_{ct,\epsilon} &= -T_7 V_{S^3}\int d\Omega_4 \left(L_1+L_2+L_f\right).
\end{align}
and the renormalised action is $S_{ren} = \lim_{\epsilon\to 0} S_{sub}$.     The chiral condensate is given by \cite{Karch:2005ms}
\begin{equation}\label{eqn:chiral}
\langle O_\phi \rangle = \lim_{\epsilon\to 0}\frac{1}{\epsilon^3 \sqrt{\mbox{det}\,g_\epsilon}}\frac{\delta S_{sub}}{\delta \phi_\epsilon} =   -2\phi_2 + \frac{\phi_0^3}{3} + \frac{R_0}{6}\phi_0 \ln{\phi_0}.
\end{equation}
If we use $m$ to denote the coefficient $\phi_0$, the flavour mass is given by the standard normalisation \cite{Kruczenski:2003be}
\begin{equation}
M_f = m \frac{\sqrt{\lambda}}{2\pi},
\end{equation}
where $\lambda$ is the t'Hooft coupling, proportional to the square of the string tension.      Although we will use $m$ to refer to the flavour mass in the rest of this paper, this relation should be kept in mind.\footnote{The mass is proportional to the asymptotic separation of the D7 branes and the D3 branes that support the $AdS_5\times S^5$.}

We look for smooth solutions of the embedding equation (\ref{eqn:embedding ode}), by which we mean that the sphere caps off smoothly, without a conical deficit.    There are two possibilities: either the $S^4$ caps off first or the $S^3$ caps off first, and thus we have two topologically distinct one-parameter families of embedding:   
\begin{description}
\item[Ungapped solutions ($S^4$ caps off first):]
D7 branes that extend all the way to the centre of AdS i.e. the angle $\psi$ is non-zero for all $0 \leq z \leq 2l$.  
\item[Gapped solutions ($S^3$ caps off first):]
D7 branes that cap off before they reach the centre i.e. the angle $\psi$ gets to $0$ for some $0<z_0 < 2 l$.   
\end{description}
For small $m l$, we expect to see only the first solution, whereas for large $m l$ only the second solution should exist.   We will see that the two branches join at some critical $m_* l$ in a continuous phase transition.    This is what happens in probe D7 brane embeddings in thermal AdS \cite{Karch:2006bv} and in the supersymmetric version of the $S^4$-sliced AdS embeddings \cite{Karch:2015kfa}.   

\subsection{Ungapped phase} 
For the first type of solution, where the $S^4$ ends at the origin $z=2 l$, we demand smoothness at that point, i.e., no conical deficit.   Looking at the first two terms in the induced metric (\ref{eqn:induced metric}) we see this requires the derivative $\psi'$ to vanish as $z\to 2l$.  Numerically we look for solutions of (\ref{eqn:embedding ode}) by integrating from the origin starting with the boundary conditions
\begin{equation}
\psi(2l-\epsilon) = \psi_0,\qquad \psi'(2l-\epsilon) = 0,\qquad 0<\psi_0\leq \pi/2 ,
\end{equation}
towards smaller $z$ until we reach the AdS boundary.  This gives a set of solutions parametrized by $\psi_0$.   There is a trivial constant solution $\psi = \pi/2$ which is the massless embedding.     

Perturbatively we can expand about the massless embedding in powers of a small dimensionless parameter $\mu l$, taking 
\begin{equation}
\cos{\psi} = (\mu l) Y_1 (z) + (\mu l)^3 Y_3(z)+... ,
\end{equation}
and solving order by order.   The leading order solution smooth at the origin is
\begin{equation}\label{eqn:perturb}
Y_1 = \frac{4 z\left(16+16 (z/l)^2 \ln{(z/2l)}-(z/l)^4\right)}{(4-(z/l)^2)^3},
\end{equation}
with the asymptotics
\begin{equation}
\cos{\psi}|_{z\to 0} = \mu z,\qquad \cos{\psi}|_{z\to 2l} = \frac{2 \mu l}{3}.
\end{equation}
Thus at this order, $\mu$ is equal to the flavour mass: $m l = \mu l + \mathcal{O}((\mu l)^3)$.\footnote{Curiously the functional combination we see in $Y_1$ also appears in the exact solution for the slipping mode in the supersymmetric version recently studied in \cite{Karch:2015vra,Karch:2015kfa}.   In fact to linear order in $m$ our solution agrees with their solution for the slipping mode.}  

\subsection{Gapped phase} 
For the second type of solution we require smoothness at the point $z_0$ where the $S^3$ caps off, i.e., absence of a conical deficit.   Looking at the first and third terms in the induced metric (\ref{eqn:induced metric}) we see that this requires the derivative $\psi'$ to diverge as $z\to z_0$.    This can be implemented numerically by starting with the boundary conditions
\begin{equation}
\psi(z_0)=\epsilon,\qquad \psi'(z_0)= -\frac{1}{\epsilon},\qquad 0<z_0<2l,
\end{equation}
and integrating towards smaller $z$.    This gives a set of solutions parametrized by $z_0$.  

Perturbatively in the large mass regime we can solve the equation order by order in inverse powers of $l/L$, taking
\begin{equation}
\cos{\psi} = X_0(z) + \frac{L^2}{l^2}X_2(z) + ...
\end{equation}
Requiring $\psi$ to vanish at $z_0=1/\mu$ where $\mu$ is a mass parameter which, as we will see in a moment, equals the flavour mass to leading order, we find the first two terms are\footnote{A very similar example of such an expansion appears when D7 branes are embedded in thermal AdS \cite{Karch:2006bv}.}
\begin{equation}\label{eqn:perturb large}
X_0(z) = \mu z,\qquad X_2(z) = \frac{\mu z-\mu^3 z^3 + 2\mu^3 z^3 \ln{(\mu z)}}{2\mu^2 L^2 (1-\mu^2 z^2)}.
\end{equation}
As expected the leading order solution is the Poincar\'e AdS embedding found in \cite{Karch:2002sh}.      The factor $1/\mu^2 L^2$ in the first term combines with $L^2/l^2$ so corrections to the leading order can be regarded as a series in inverse powers of $(\mu l)^2$.    Expanding near $z=0$ we find that the flavour mass $m l  = \mu l + \mathcal{O}(1/\mu l)$.

\begin{figure}
\centering
\includegraphics[width=6in]{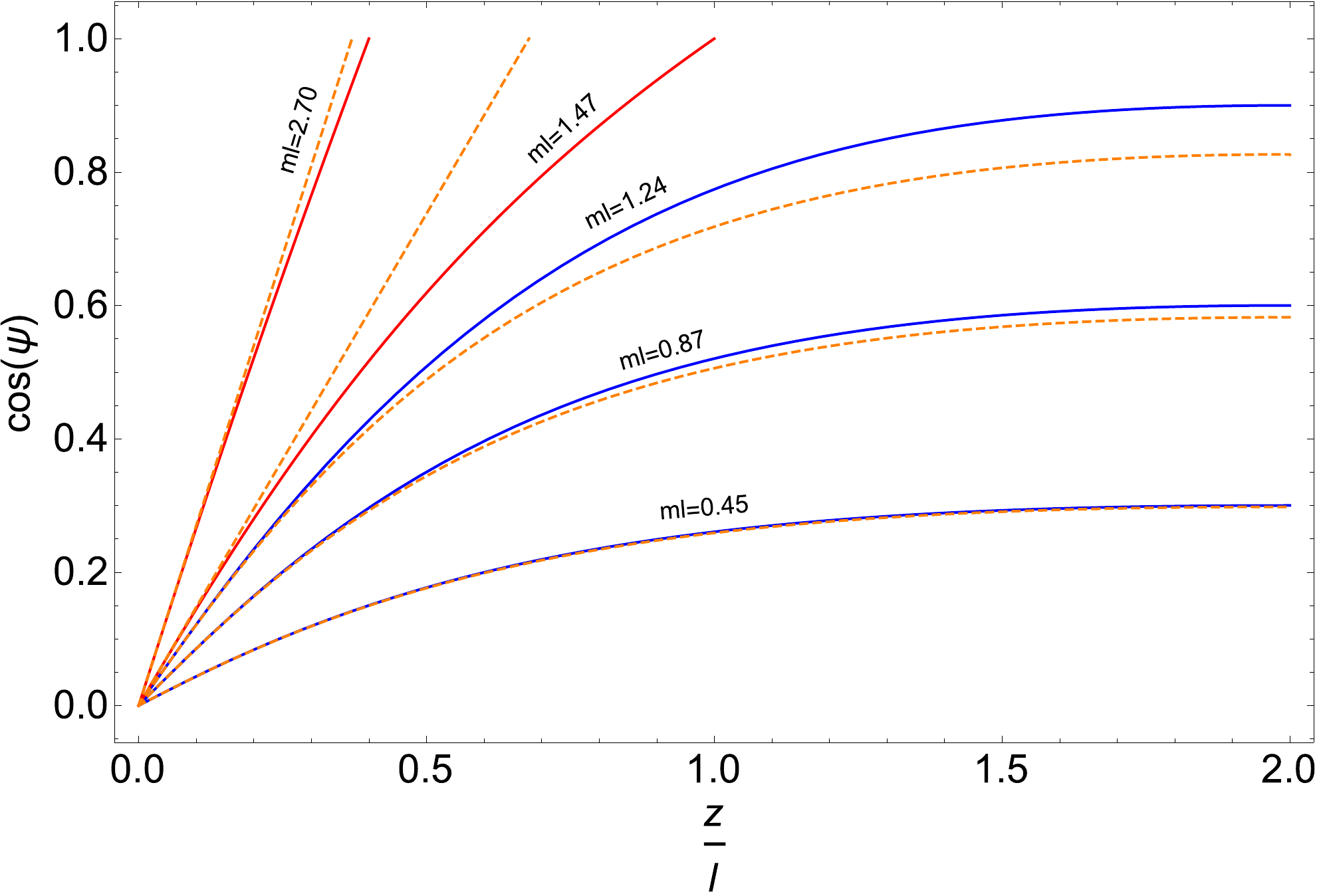}
\caption{The slipping mode $\cos{\psi}$ of numerical probe D7 brane embeddings in $S^4$-sliced $AdS_5\times S^5$ for five different masses $m l$.   The lower three masses (solid blue lines) correspond to ungapped solutions and were obtained with the starting values $\psi_0=0.3$, $\psi_0=0.6$ and $\psi_0=0.9$ respectively.   The upper two masses (solid red lines) correspond to gapped solutions and were obtained using $z_0/l=1$ and $z_0/l=0.4$ respectively.   The dashed orange lines represent leading order perturbative approximations $Y_1$ and $X_0$ respectively. }
\label{fig:ne}
\end{figure}

\begin{figure}
\centering
\includegraphics[width=6in]{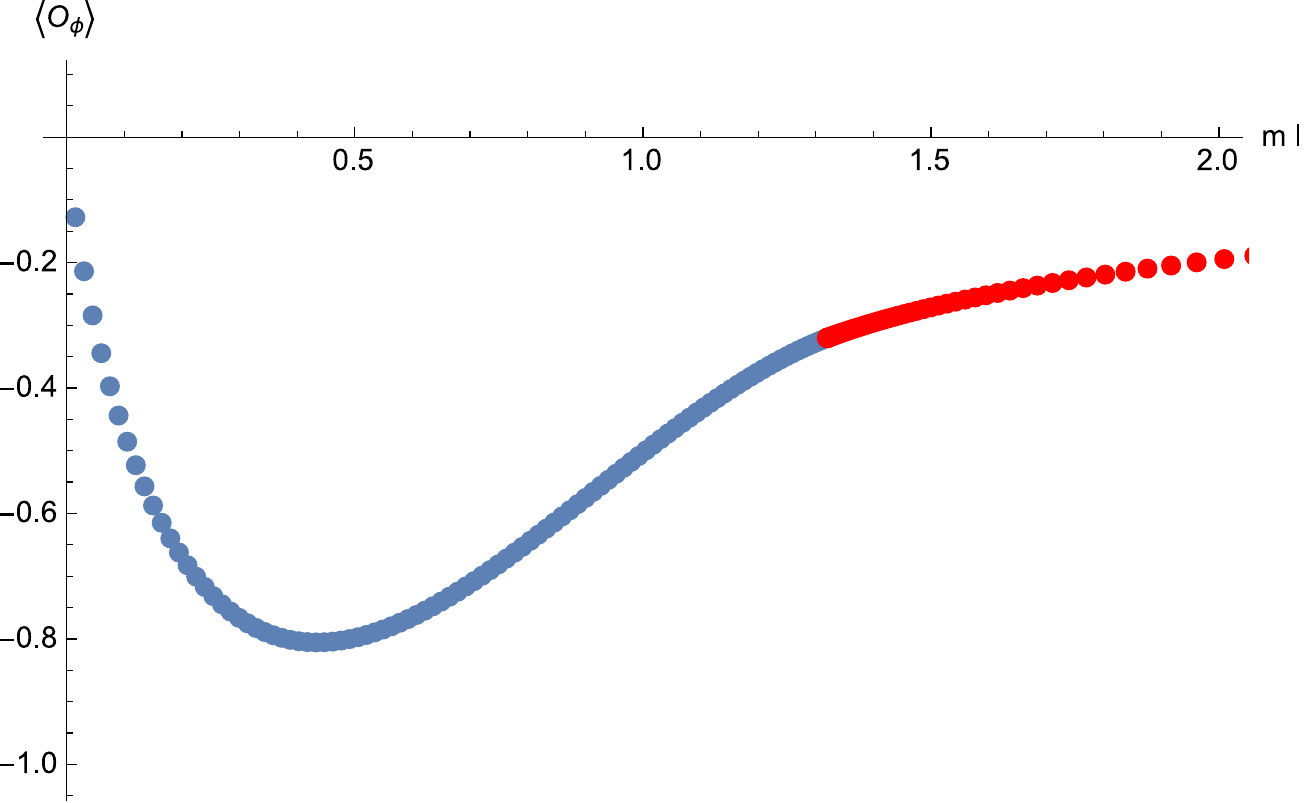}
\caption{Chiral condensate $\langle O_\phi \rangle$ for the embeddings in Fig. \ref{fig:ne} as a function of the mass $m l$.   The blue line represents the ungapped phase, the red line corresponds to the gapped phase.   The two branches merge at $m_* l \approx 1.32$ in a continuous phase transition.  Note that $\langle O_\phi \rangle \to 0$ as $ml\to\infty$.}
\label{fig:ne1}
\end{figure}

\subsection{Numerical solutions}  
In Fig. \ref{fig:ne} we show numerical embeddings for five different values of $ml$, as well as the leading order perturbative approximations $Y_1$ and $X_0$ from (\ref{eqn:perturb}) and (\ref{eqn:perturb large}).\footnote{Numerically we found it easier to work with the differential equation for $y=\cos{\psi}$.}  It shows that the approximations are quite accurate for small $m l \leq 0.5$ and large $m l \geq 3$ respectively.   

We found no ungapped solutions for $m l \gtrsim 1.32$ and no gapped solutions for $m l \lesssim 1.32$.     We conclude that the critical mass at the transition is $m_* l  \approx 1.32$ in this example.   The chiral condensate (\ref{eqn:chiral}) is plotted in Fig. \ref{fig:ne1} and backs up this conclusion as it shows the two branches joining at $m l=m_* l$.   It indicates the phase transition is continuous.       

We note that our $m_* l$ is greater than the critical mass $m_* l = 1$ in the supersymmetric version of these embeddings \cite{Karch:2015kfa}.

\subsection{Continuation to de Sitter}\label{subsec:continue}
Continuing the brane embeddings to the $dS_4$ slicing of $AdS_5\times S^5$ is as simple as Wick rotating the polar angle on the sphere $S^4$.   The induced metric (\ref{eqn:induced metric}) becomes
\begin{equation}
\gamma = L^2\left(\frac{1}{z^2}+\psi'^2(z)\right)dz^2 +  \frac{L^2}{z^2}f(z) \,\left(-d\tau^2 + l^2 \cosh^2{(\tau/l)} \,d\Omega_{3}^2\right)  + L^2\sin^2{\psi(z)}\,d\Omega_3^2.
\end{equation}
There is a crucial difference between ungapped solutions and gapped solutions.     In Euclidean signature the D7 brane in the ungapped solution ends where the $S^4$ caps off.   In Lorentzian signature this point is a horizon through which we can continue into an FRW geometry, as explained in Sec. \ref{subsec:horizon}.  On the other hand the D7 brane in the gapped solution ends where the $S^3$ caps off.   The analytically continued Lorentzian embedding also ends there.    Thus ungapped solutions have an FRW geometry living on their Lorentzian worldvolume, whereas gapped solutions do not.  

To compute the EE in the next section we continue to use the Euclidean embeddings for computational ease (the method is defined in Euclidean signature), however we will see from the results that the EE is in fact probing the Lorentzian solutions.

\section{Flavour entanglement entropy} 
\label{sec:flavourEE}
We compute the flavour EE of a spherical region in de Sitter using the probe brane embeddings obtained in Sec. \ref{sec:flavours}.     We focus on the EE in global de Sitter since we know from Sec. \ref{sec:cft} that it contains static de Sitter as a special case.         

\subsection{Method} 
To avoid computing the backreaction of the probe branes on the geometry, we apply the method of \cite{Karch:2014ufa}.    Inspired by \cite{Lewkowycz:2013nqa} they argued that the correction to the EE at linear order in the strength of the backreaction (a combination of the brane tension and Newton's constant) can be computed by evaluating the variation in the probe brane action ((\ref{eqn:dbi}) plus counterterms) with respect to changes in the induced metric with respect to a number $n$.   The $n$ refers to certain ``replica'' geometries which are the bulk extensions of the $n$-fold covers of the boundary which would be needed to compute the EE directly in the field theory using the replica trick.    In general, these bulk replica geometries are hard to find.   For the special case where the entangling surface is a sphere, they can be obtained by mapping the induced metric on the brane to the (Euclidean) hyperbolic slicing (\ref{eqn:hyperbolic metric}) of AdS and changing the periodicity of the Euclidean time coordinate from $2\pi L$ to $2\pi n L$.     Due to the fact that these geometries are deformations of the original $n=1$ geometry along an on-shell path, the result for the flavour contribution to the EE requires only the brane embedding for the $n=1$ geometry.   As the procedure in \cite{Karch:2014ufa} is technically quite involved we now outline how it works in our case.    

In the absence of probe branes, we established in Sec. \ref{sec:cft} that the EE of a spherical region on the de Sitter boundary is equal to the thermal entropy on $\mathbb{R}\times H^{3}$ which is proportional to the area of the horizon in the hyperbolic slicing of AdS.    With probe D7 branes added, we can map the correction to the EE of the spherical region to the correction to the thermal entropy on $\mathbb{R}\times H^{3}$.   To calculate the latter we do the following \cite{Karch:2014ufa}\footnote{We work in Euclidean signature.  The Euclidean hyperbolic slicing of $AdS_5$ is the hyperbolic slicing (\ref{eqn:hyperbolic metric}) with Euclidean time $\hat{\sigma} = -i\sigma$.}

\begin{enumerate}
\item
Transform the induced metric to the hyperbolic slicing.    Our Euclidean induced metric (\ref{eqn:induced metric}) is $\gamma = g_{AdS_5} + L^2 \psi'^2(z) dz^2 + L^2 \sin^2{\psi(z)}\,d\Omega_3^2$, where $g_{AdS_5}$ is in the $S^4$-slicing.   There is a set of transformations between the $S^4$-slicing coordinates $(z,\Omega_4)$ and the hyperbolic slicing coordinates $(\hat{\sigma},\rho,u,\Omega_2)$.    The relevant part is the transformation of the $z$ coordinate which takes the form $z(\hat{\sigma},\rho,u)$.   So for the induced metric we get 
\begin{equation}\label{eqn:induced metric hyperbolic}
\gamma = g_{AdS_5}^{hyp} +  L^2 \psi'^2(z(\hat{\sigma},\rho,u)) dz(\hat{\sigma},\rho,u)^2 + L^2 \sin^2{\psi(z(\hat{\sigma},\rho,u))}\,d\Omega_3^2,
\end{equation}
where $g_{AdS_5}^{hyp}$ denotes (\ref{eqn:hyperbolic metric}) with Euclidean time $\hat{\sigma}$ and $dz(\hat{\sigma},\rho,u)$ means $(\partial_{\hat{\sigma}}z) d\hat{\sigma}+(\partial_\rho z) d\rho + (\partial_u z) du$. 

\item	
Write down the induced metric $\gamma_n$ for the replica geometries where the integer $n=1,2,3...$ is the replica number.   This is given by
\begin{align}\label{eqn:replica metric}
\gamma_n &= g_{AdS_5}^{hyp,n} +  L^2 \psi'^2(z(\hat{\sigma},\rho,u)) dz(\hat{\sigma},\rho,u)^2 + L^2 \sin^2{\psi(z(\hat{\sigma},\rho,u))}\,d\Omega_3^2 \nonumber \\
g_{AdS_5}^{hyp,n} &= F_n(\rho)d\hat{\sigma}^2 + \frac{d\rho^2}{F_n(\rho)}+\rho^2\left(du^2+\sinh^2{u}\,d\Omega_2^2\right) \nonumber \\
F_n(\rho) &= \frac{\rho^2}{L^2}-1-\frac{\rho_h^2(n)}{\rho^2}\left(\frac{\rho_h^2(n)}{L^2}-1\right),\qquad \rho_h(n) = \frac{L}{4 n}\left(1+\sqrt{1+8n^2}\right).
\end{align}
These metrics are just the hyperbolic AdS black holes first discussed in \cite{Emparan:1999gf}.   The dependence of $\rho_h$ on $n$ is designed to ensure that for the $n$-th replica the Euclidean time $\hat{\sigma}$ is periodic with period $2\pi n L$ with no conical singularity.    Note that $\gamma_1=\gamma$ is the induced metric (\ref{eqn:induced metric hyperbolic}) in the $n=1$ geometry.   
\item
Continuing $n$ to non-integer values, the correction to the EE is given by the first variation away from $n=1$ of the brane action evaluated on the replica geometries \cite{Karch:2014ufa}
\begin{align}\label{eqn:eecorrection}
\mathcal{S}_{EE}^{(1)} &=  \delta_n S_{brane} + \delta_n S_{ct} \nonumber \\
& = \delta\rho_h\, T_7 \int_{\rho=\rho_h} d^7 y \sqrt{\gamma} - T_7 \int_{\rho_h}^{\rho_\epsilon} d\rho \int d^7 y\, \partial_n \sqrt{\gamma_n}|_{n=1} + \partial_n S_{ct,brane}|_{n=1},
\end{align}
where $\delta\rho_h \equiv \rho_h'(1)=-L/3$, the symbol $\rho_h$ denotes the horizon in the $n=1$ geometry which is just $L$, and $\rho_\epsilon$ is a regulator that we will take to infinity (see below).   The integration $d^7 y$ is over the coordinates $(u,\hat{\sigma},\Omega_2,\Omega_3)$, where $\hat{\sigma}$ is integrated over $[0,2\pi L)$.   
\item
Counterterms are dealt with as follows.    The brane action in the original coordinates had divergences at small $z$ which were regulated by the counterterms (\ref{eqn:counterterms}).    The small $z$ divergences map to large $u$ and large $\rho$ divergences.   As in \cite{Karch:2014ufa} we use a partial renormalisation whereby we do not cancel the large $u$ divergences in order to see the UV divergences of the EE.  We use a cut-off $u_{max}$ which is the same one used to compute the CFT EE from empty AdS (\ref{eqn:dscutoff}).        We do subtract off the large $\rho$ divergences, however we will find that the second integral in (\ref{eqn:eecorrection}) is convergent at large $\rho$ and that the variation of all the counterterms with $n$ vanishes as $\rho\to\infty$ with the exception of the volume counterterm which leaves behind a finite piece.   

\end{enumerate}

We will compute the EE of a spherical region of angular size $\theta_0$ at time $\tau_0$ in global de Sitter.   We expect the result to be a function only of the combination $R = l\sin{\theta_0}\cosh{(\tau_0/l)}$ and when $R<l$ to concur with the analogous question posed in static de Sitter.    As shown in Sec. \ref{sec:globaldesitterfoliation} to map this EE to the thermal entropy of $\mathbb{R}\times H^3$ we need the hyperbolic foliation (\ref{eqn:boosted hyperbolic slicing}) with boost $\beta$ and rotation $\alpha$ which are related to $\tau_0,\theta_0$ by (\ref{eqn:globaldsintersection}).    The relevant part of the coordinate transformation $z(\hat{\sigma},\rho,u)$ is given by (\ref{eqn:observe global})\footnote{The $i$ in this comes from the fact that for non-zero $\tau_0$ this transformation is real only in Lorentzian signature ($\sigma = i\hat{\sigma}$) but since the integral (\ref{eqn:eecorrection}) uses Euclidean time, it is convenient to continue working in Euclidean signature and carry the $i$.}   
\begin{align}\label{eqn:ztransf}
& z(\hat{\sigma},\rho,u) = \frac{2l}{L}\left(y_{-1}-\sqrt{y_{-1}^2-L^2}\right), \nonumber \\ 
& y_{-1} = \sech{(\tau_0/l)}\left( \csc{\theta_0}\,\rho\cosh{u} + \cot{\theta_0}\,\sqrt{\rho^2-L^2} \cos{(\hat{\sigma}/L)}\right) - \nonumber\\
&\qquad\qquad\qquad\qquad\qquad\qquad -i \tanh{(\tau_0/l)}\sqrt{\rho^2-L^2}\sin{(\hat{\sigma}/L)}.
\end{align}
We now express the metric determinant and its variation that we need to compute the EE correction (\ref{eqn:eecorrection}) in terms of the slipping mode of the embedding and the transformation function (\ref{eqn:ztransf}).      Details of the calculation can be found in Appendix \ref{app:metric}.  The result is 
\begin{align}\label{eqn:det}
& \sqrt{\gamma} = L^3 \sqrt{g_{S^2}}\,  \sqrt{g_{S^3}}\,  \rho^3 \sinh^2{u}\, \sin^3{\psi} \sqrt{1+z^2 \psi'^2}\nonumber \\
&\partial_n \sqrt{\gamma_n}|_{n=1} = \frac{\sqrt{\gamma}\, L^2 \psi_z^2}{1+z^2\psi_z^2}\frac{\left(\frac{\rho^2}{L^2}-1\right)^2(\partial_\rho z)^2 - (\partial_{\hat{\sigma}} z)^2}{\frac{3\rho^2}{L^2}\left(\frac{\rho^2}{L^2}-1\right)^2},
\end{align}
where $\psi_z = \psi'(z)$.   

\paragraph{Counterterms}
As discussed above we add counterterms only on the large $\rho=\rho_\epsilon$ surface.   The AdS part of the induced metric on this regulator surface is
\begin{equation}
g_\epsilon = F_n(\rho_\epsilon) d\hat{\sigma}^2 + \rho_\epsilon^2\left(du^2 + \sinh^2{u}\,d\Omega_2^2\right).
\end{equation}
Since $\partial_n F_n(\rho_\epsilon)|_{n=1}  = \frac{2}{3}L^2 \rho_\epsilon^{-2}$ the contributions from most counterterms will vanish in the limit $\rho_\epsilon\to\infty$.     The argument is identical to the Poincar\'e AdS case \cite{Karch:2014ufa}.    Namely because $\partial_n \sqrt{\mbox{det}\,g_\epsilon}|_{n=1}=\mathcal{O}(\rho_\epsilon^0)$ the volume counterterm contributes.  The Ricci scalar $R_\epsilon \sim \rho_\epsilon^{-2}$, the curvature scalar in the logarithmic term vanishes, and the slipping mode decays as $\phi \sim z \sim \rho_\epsilon^{-1}$, therefore the contributions from all other counterterms vanish. The contribution to the EE from the volume counterterm is
\begin{equation}\label{eqn:countertermvariation}
 \partial_n S_{ct,brane}|_{n=1}  = \frac{\pi}{6} V_{S^2}V_{S^3} L^8 T_7 \int_0^{u_\infty} du\, \sinh^2{u} .
\end{equation}

\paragraph{Strategy}
In principle it is possible to insert the numerical brane embeddings we obtained into (\ref{eqn:det}) and compute the integrals numerically, however we choose not to go down this route, and instead use the leading order perturbative embeddings to examine the behaviour of the EE correction (i) in the small mass regime $m l \ll 1$.  (ii) in the large mass regime $m l \gg 1$.   In both limits the size of the entangling region in units of de Sitter length, $R/l$, can take arbitrary values.

\subsection{Small mass regime $m l \ll 1$}
If we take the leading order perturbative solution (\ref{eqn:perturb}) describing the ungapped phase and insert it into (\ref{eqn:det}), the determinant of the metric and its variation are given to order $(m l)^2$ by
\begin{align}
\sqrt{\gamma} &=  (\rho L)^3 \sinh^2{u}\, \sqrt{g_{S^2}}\, \sqrt{g_{S^3}} \left( 1-\frac{(m l)^2}{2} \left(3 Y_1^2-z^2 Y_{1z}^2\right)\right)  \nonumber \\ 
\partial_n \sqrt{\gamma}|_{n=1} &= \left(m l L\, Y_{1z}\right)^2 \rho L^5 \sinh^2{u}\, \sqrt{g_{S^2}}\, \sqrt{g_{S^3}}  \frac{\left(\frac{\rho^2}{L^2}-1\right)^2 (\partial_\rho z)^2 -(\partial_{\hat{\sigma}} z)^2}{3 \left(\frac{\rho^2}{L^2}-1\right)^2}.
\end{align}
where $Y_{1z}=Y_1'(z)$.   Taking into account the counterterms (\ref{eqn:countertermvariation}), the expression (\ref{eqn:eecorrection}) for the flavour contribution to the EE in de Sitter becomes
\begin{equation}\label{eqn:eeintegralsum}
\mathcal{S}_{EE}^{(1)} = \left(-\frac{2\pi}{3}V_{S^2}V_{S^3}L^8 T_7\right)\left(I_{horizon}+I_{bulk}+I_{ct}\right),
\end{equation}
where, after substituting $\rho = L \cosh{\zeta}$,
\begin{align}
& I_{horizon} = \int_0^{u_{\infty}}du\,\sinh^2{u}\left(1-\frac{(m l)^2}{2}\left(3Y_1^2-z^2 Y_{1z}^2\right)\right)\nonumber \\
& I_{bulk} = (m l)^2 \int_0^{u_{\infty}}du\int_0^\infty d\zeta \int_0^{2\pi L} \frac{d\hat{\sigma}}{2\pi L} Y_{1z}^2 \frac{\cosh{\zeta}}{\sinh^3{\zeta}}\,\sinh^2{u} \left(\sinh^2{\zeta}(\partial_\zeta z)^2-L^2(\partial_{\hat{\sigma}} z)^2\right) \nonumber \\
& I_{ct} = -\frac{1}{4} \int_0^{u_{\infty}}du\,\sinh^2{u}.
\end{align}
In the horizon term we use $z$ evaluated at the horizon of the $n=1$ geometry $\rho=L$, whereas in the bulk term we need the full transformation (\ref{eqn:ztransf}).     The regulator in the $u$ integral is the same as (\ref{eqn:dscutoff})
\begin{equation}
\cosh{u_\infty} = \frac{R}{\epsilon}\left(1+\frac{\epsilon^2}{4 l^2}\right),\qquad R = l \sin{\theta_0}\cosh{(\tau_0/l)}.
\end{equation}
We can evaluate the integrals for the massless embedding ($m=0$) and separate out the massive contribution 
\begin{equation}\label{eqn:correction final}
\mathcal{S}_{EE}^{(1)} = \frac{t_0 L^3 V_{S^2}}{128 G_N}\left(-\frac{2 R^2}{\epsilon^2}+2\ln{(2 R/\epsilon)}-\frac{R^2}{l^2}+ 1 + \mathcal{S}(m l,R/l,\epsilon/l) \right),
\end{equation}
where the coefficient at the front has been expressed terms of the dimensionless backreaction parameter $t_0 = 16\pi G_N L^{-3} T_0$ where $T_0 = V_{S^3} T_7 L^8$ to compare to the CFT result before we added flavours and to the flat space result in \cite{Karch:2014ufa}.   Note that the EE correction for massless flavours is precisely (\ref{eqn:eecft4}) times $-t_0/8$ which was also the case in flat space \cite{Karch:2014ufa}.  

The massive contribution is denoted by $\mathcal{S}$ and is equal to $(-16/3)(I_{horizon}+I_{bulk})$.  It has an additional UV log divergence and a finite piece.     We can get the divergence by expanding the integrands at large $u$.    We find\footnote{The divergent piece has a factor of $-16/3$ from the horizon and a factor of $8/3$ from the bulk.   We could also have written it in terms of $\ln{(\epsilon/2 R)}$ but we choose to absorb the $\ln{(2R/l)}$ part into the finite piece.}
\begin{equation}\label{eqn:j}
\mathcal{S} = -\frac{8  }{3}(m R)^2 \ln{(\epsilon/l)} + \mathcal{S}_{finite}(m l, R/l).
\end{equation}
We compute the finite piece $\mathcal{S}_{finite}$ numerically.   Note that the horizon integral is manifestly a function only of the combination $R/l =  \sin{\theta_0}\cosh{(\tau_0/l)}$.     In the bulk integral this is not obvious because the full transformation (\ref{eqn:ztransf}) depends on $R$ and another combination, for example $K=\sin{\theta_0}$.  However, we find when we expand the integrand at large $u$ and average over one cycle of $\hat{\sigma}$ the expression is independent of the value of $K$ at every order.  This is also seen numerically.\footnote{We have verified it only up to second order in the large $u$ expansion but the numerical evidence is convincing.  There must be a nice way to prove it but we have not been able to find it.}

We briefly review the analogous result in Minkowski space, eqn. 55 of \cite{Karch:2014ufa} in order to make some comparisons.   The de Sitter length scale $l$ does not exist in flat space, and we have only three dimensionless combinations of the flavour mass $m$, the sphere size $R$ and the UV cut-off $\epsilon$.   In our notation, the result of \cite{Karch:2014ufa} takes the form (\ref{eqn:correction final}) without the $R^2/l^2$ piece and with the massive part given by the exact expression
\begin{equation}\label{eqn:flatspace}
\mathcal{S}_{flat} =  -\frac{8  }{3}(m R)^2 \ln{(\epsilon/2 R)} -\frac{16}{9}m^2 R^2 + \frac{4}{45} m^4 R^4,\qquad m R <1.
\end{equation}
The first part of (\ref{eqn:flatspace}) contains a UV log divergence of the same form as in (\ref{eqn:j}), with the same coefficient.     It must be related to the universal correction to the EE of the form $(m R)^2 \ln{(m \epsilon)}$, discussed in Sec. \ref{sec:setup}, expected to appear when 4d field theories are deformed by a relevant operator.   We will have more to say about this universal term in the context of de Sitter.   The finite part of the flat space result is scheme and state dependent, however, since it originates from the same type of calculation as we have done in de Sitter, it will be useful as a consistency check.

\begin{figure}
\centering
\includegraphics[width=3in]{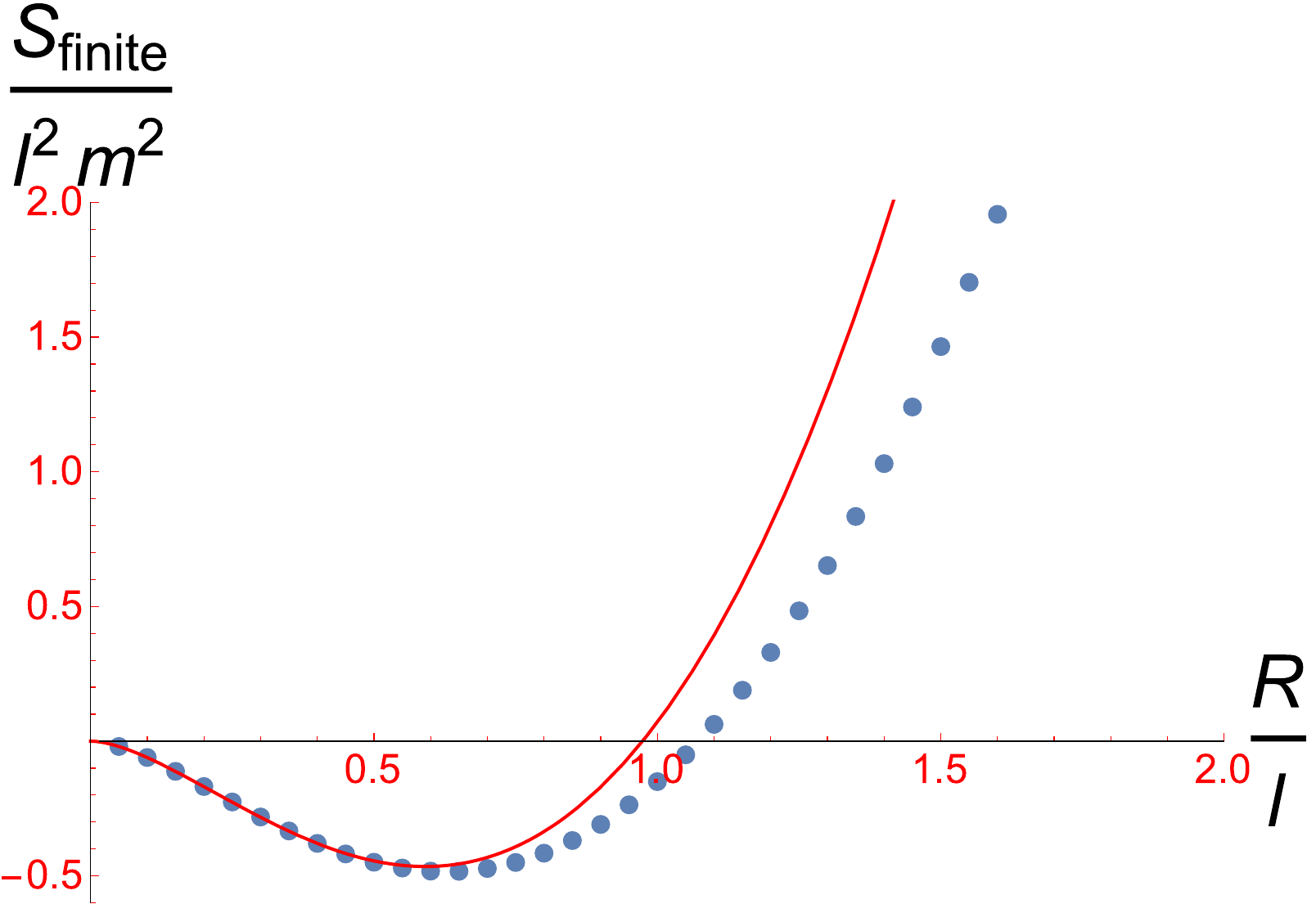}\,\,\includegraphics[width=3in]{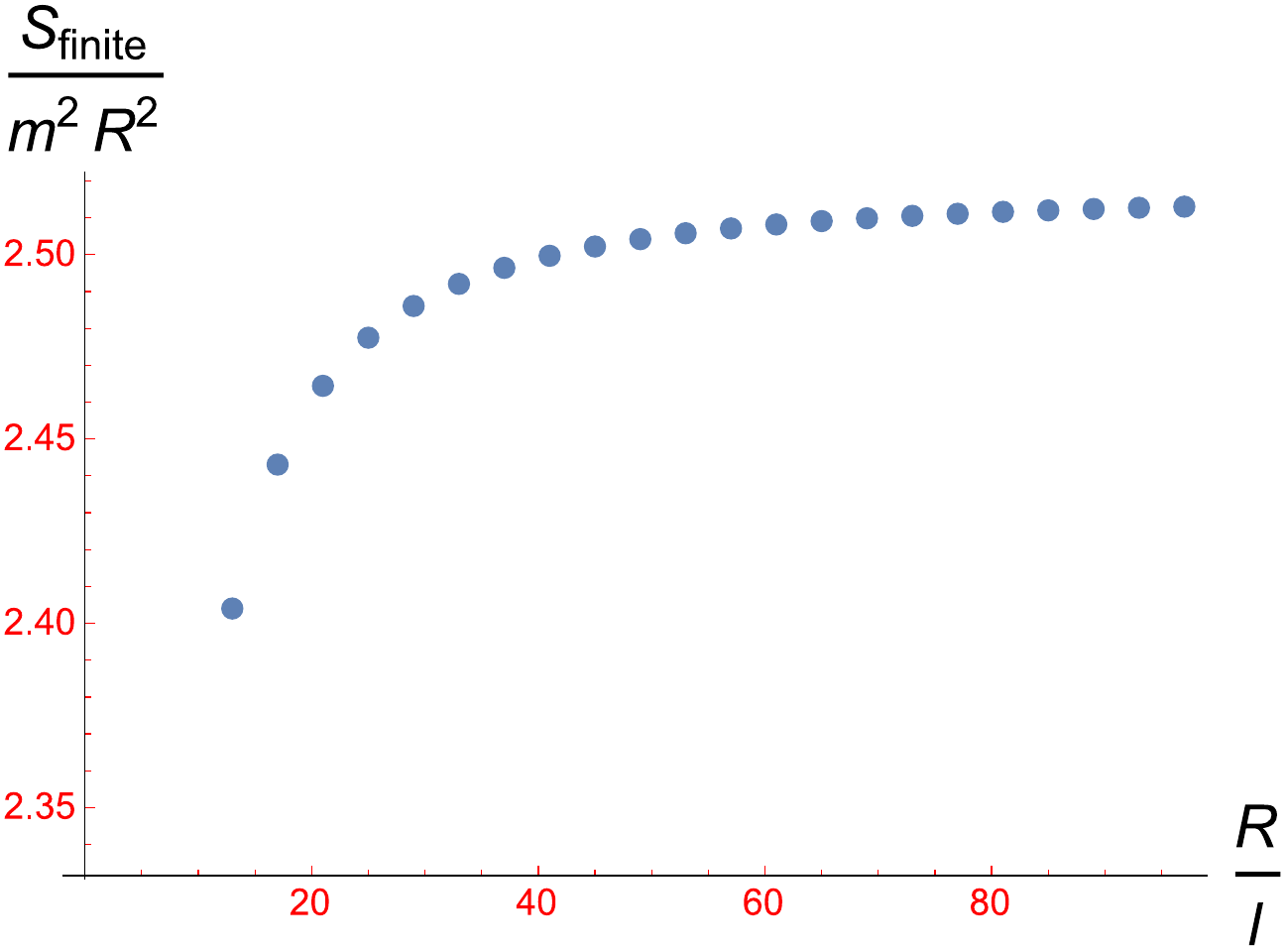}
\caption{Small mass regime $m l\ll 1$.  On the left is the finite part of the massive correction to the EE, $\mathcal{S}_{finite}$ per unit mass squared, as a function of the proper size $R/l$ of the entangling sphere.   The continuous red curve is the corresponding correction in flat space.   On the right is the behaviour of $ \mathcal{S}_{finite}/m^2 R^2$ for very large spheres $R/l\gg 1$ which is consistent with the asymptotics $\mathcal{S}_{finite} \simeq a_5 (R/l)^2+ a_6 \ln{(R/l)}$.}
\label{fig:ne2}
\end{figure}

\noindent In Fig. \ref{fig:ne2} we show the massive EE correction $\mathcal{S}_{finite}$ in de Sitter space for a range of values of $R/l$ at fixed small mass $m l \ll 1$.   We note that in this approximation $\mathcal{S}_{finite}$ scales uniformly as a function of $ml$: $\mathcal{S}_{finite}/(m l)^2$ is independent of $m l$.     

In particular, we do not see a term logarithmic in the mass, as might be expected to appear from a universal $(m R)^2 \ln{(m \epsilon)}$ term, paired with the $(m R)^2 \ln{(\epsilon/l)}$ divergent term which we do see.    Note that although we are in the $m l \ll 1$ regime, the value $R/l$ is unrestricted, so $mR$ can be arbitrarily large.   However, we will see this logarithmic term appear in the opposite limit $ml \gg 1$.

From Sec. \ref{sec:setup} we expect that for small spheres, $R/l \ll 1$, the EE in de Sitter behaves similar to flat space.    This can be checked by taking the flat space result (\ref{eqn:flatspace}) and using the extra scale $l$ to separate it into a divergent part and a finite part.  We get
\begin{equation}
\mathcal{S}_{flat,finite} = \frac{8}{3}(m R)^2 \ln{(2R/l)} - \frac{16}{9}(m R)^2.
\end{equation}
We see from Fig. \ref{fig:ne2} that this is a good fit for small $R/l$ that $\mathcal{S}_{finite}$ only starts to deviate appreciably from when $R/l \approx 1$.    

For very large spheres i.e. at late times, $R/l \gg 1$, and from Sec. \ref{sec:setup} we expect $\mathcal{S}_{finite}\simeq a_5 (R/l)^2+a_6 \ln{(R/l)}$.     The plot in the right side of Fig. \ref{fig:ne2} is consistent with this asymptotics.    A best fit estimate gives $a_5 \approx 2.5 (m l)^2$ and $a_6 \approx -7.6 (m l)^2$.   Note there are additional CFT contributions to the overall $a_5$ and $a_6$.     A non-zero $a_6$ in this limit is consistent with the arguments in \cite{maldacenapimentel}.   This follows from the discussions in Sec. \ref{sec:setup}, Sec. \ref{subsec:horizon} and Sec. \ref{subsec:continue}.  An interesting consistency check would be to compute the maximum value of the scale factor in the FRW region on the brane in the backreacted geometry and compare it to our net (CFT+massive) $a_6$.     

\subsection{Large mass regime $ml \gg 1$} 
We take the leading order perturbative solution $X_0$ describing the gapped phase (\ref{eqn:perturb large}) and insert it\footnote{It is just the Poincar\'e AdS embedding but the coordinate transformation is different from the flat case.} into (\ref{eqn:det}). We end up with (\ref{eqn:eeintegralsum}) but this time with
\begin{align}
& I_{horizon} = \int_0^{u_{\infty}}du\,\sinh^2{u}\left(1-m^2 z^2\right) \nonumber \\
& I_{bulk} = m^2 \int_0^{u_{\infty}}du\int_0^\infty d\zeta \int_0^{2\pi L} \frac{d\hat{\sigma}}{2\pi L}\cdot (1-m^2 z^2)\frac{\cosh{\zeta}}{\sinh^3{\zeta}} \,\sinh^2{u} \left(\sinh^2{\zeta}(\partial_\zeta z)^2-L^2(\partial_{\hat{\sigma}} z)^2\right) \nonumber \\
& I_{ct} = -\frac{1}{4} \int_0^{u_{\infty}}du\,\sinh^2{u}.
\end{align}
Although the integrands are simpler than in the small mass case, the branes end at $z=1/\mu$ (where $\mu=m$ at this order) and so we must integrate only over the region $z<1/m$.    This is quite a non-trivial constraint, although for the horizon and counterterm integrals it translates into the requirement $\cosh{u} > m R (1+1/4(m l)^2)$.    It follows that it makes a difference only when $R \gtrsim 1/m$, however this is nearly always the case.   The constraint is not difficult to impose numerically provided $0<R/l <1$.    For $R/l>1$ the bound on the integration region in the bulk integral is complex and we have so far not found a way to implement it numerically.  

\begin{figure}
\centering
\includegraphics[width=3in]{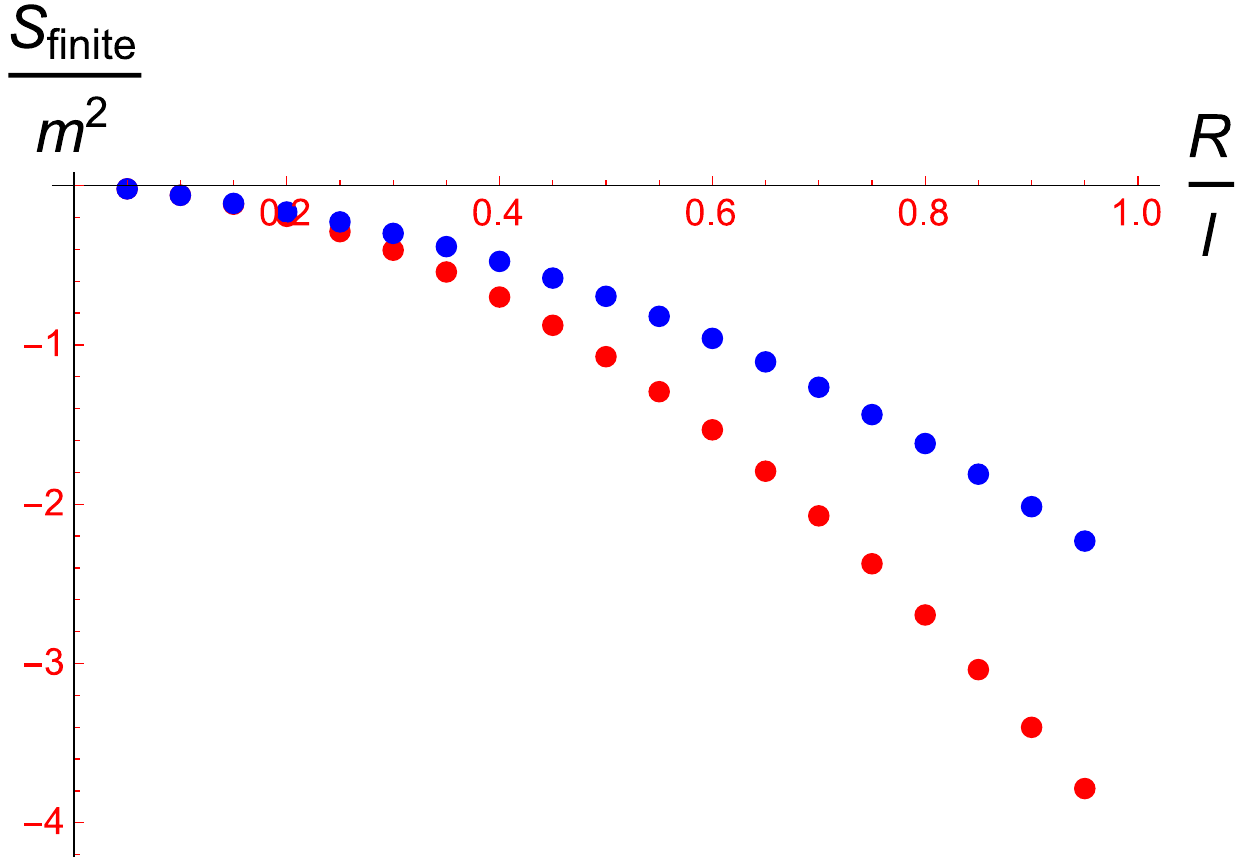}\,\,\,\,\includegraphics[width=3in]{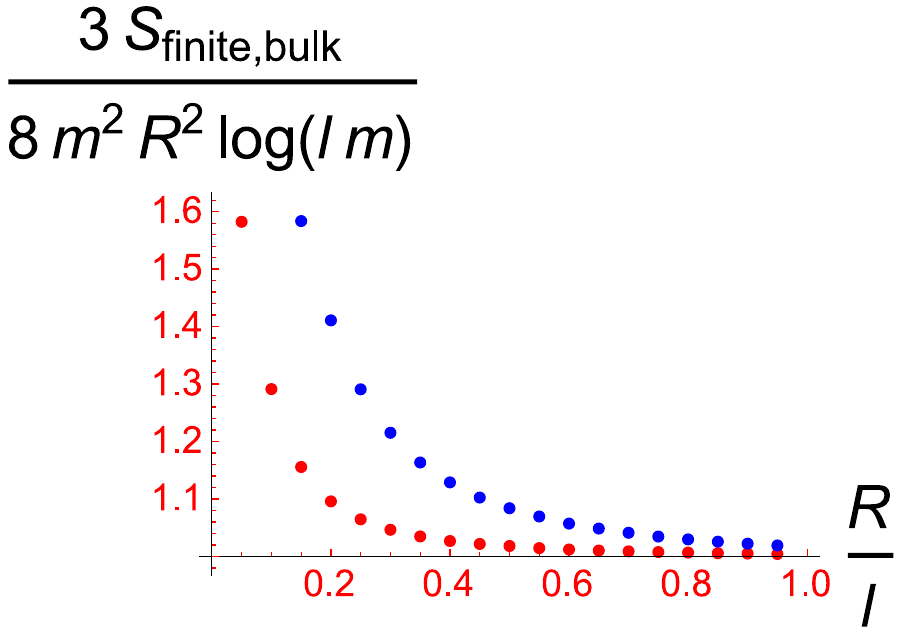}
\caption{Large mass regime $ml\gg 1$.    On the left is the finite part of the massive correction to the EE, $\mathcal{S}_{finite}$ per unit mass squared, as a function of the proper size $R/l$ of the entangling sphere, for two different masses, $m l=10$ (red) and $m l=5$ (blue).   The right is the bulk part of this correction for the same two masses, consistent with $\mathcal{S}^{finite}_{bulk} \to \frac{8}{3}m^2 R^2 \ln {(m l)}$ for large $m l$ or large $R/l$.}
\label{fig:ne3}
\end{figure}

Isolating the contribution from the $u_{max}$ limit in all three integrals, we get exactly the same mass-independent contributions as in (\ref{eqn:correction final}), and in addition the same mass-dependent UV log divergence as in (\ref{eqn:j}).      The finite part of the massive EE correction $\mathcal{S}_{finite}$ is shown in Fig. \ref{fig:ne3} for $0<R/l<1$ for two different large masses $m l =10$ and $m l = 5$.   Note that because of the non-trivial constraint the correction does not scale uniformly as a function of mass i.e. $\mathcal{S}_{finite}/(m l)^2$ depends on $m l$.   

We now argue that in the large mass regime the EE contains the universal term $(m R)^2 \ln{(m \epsilon)}$.     We separate the finite part of the massive correction into three pieces:
\begin{equation}
\mathcal{S}_{finite} = \mathcal{S}_{finite,0} + \mathcal{S}_{finite,horizon} + \mathcal{S}_{finite,bulk}.
\end{equation}
The first part $\mathcal{S}_{finite,0}$ comes from the lower limit in the $m$-independent part of the integrand (i.e. $\int du\, \sinh^2{u}$).   The expression is quite complicated but in the limit $m l \gg 1$ it is given by
\begin{equation}\label{eqn:semi1}
\mathcal{S}_{finite,0} \approx 2 m^2 R^2 - 2 \ln{ (2 mR)}.
\end{equation}
The second part $\mathcal{S}_{finite,horizon}$ comes from integrating the $m$-dependent part of the horizon integrand.    It can be done analytically, and for large $m l$ or large $R/l$ its leading order behaviour is
\begin{equation}
\mathcal{S}_{finite,horizon} \approx -\frac{16}{3}(m R)^2 \ln{(m l)}.
\end{equation}
The third part $\mathcal{S}_{finite,bulk}$ is the bulk part of the integral and we have only been able to do it numerically, and only in the range $0<R/l<1$ (for arbitrary large $ml$).     The numerics indicate that its behaviour in this range, as a function of $R/l$ and $ml$, is quite similar to $\mathcal{S}_{finite,horizon}$ and that for large $m l$ or large $R/l$ it is well approximated by
\begin{equation}
\mathcal{S}_{finite,bulk} \approx \frac{8}{3}(m R)^2 \ln{(m l)}.
\end{equation}
as shown in Fig. \ref{fig:ne3}.    We have not ruled out that its behaviour changes as $R/l$ crosses $1$: to do so we would need to figure out how to do the integral for $R/l >1$.   However we have no reason to expect a discontinuity.    It would be good to confirm this.   Adding together the three contributions to $\mathcal{S}_{finite}$, these arguments suggest that in the regime $ml \gg 1$ the massive contribution to the flavour EE in de Sitter is given by 
\begin{equation}
\mathcal{S} =  -\frac{8  }{3}(m R)^2 \ln{(\epsilon/l)} - \frac{8}{3}(m R)^2 \ln{(m l)} + \mbox{subleading} = -\frac{8  }{3}(m R)^2 \ln{(m \epsilon)} + \mbox{subleading}.
\end{equation}
provided $mR > 1$ (recall that if $ mR <1$ the bound $z<1/m$ is automatically satisfied).    This is the universal contribution identified in \cite{Hung:2011ta,Lewkowycz:2012qr}.      There is also a $\ln{(m \epsilon)}$ term without an $(mR)^2$ factor coming from (\ref{eqn:semi1}).   In fact the coefficients of both these terms agree with the $m R \gg 1$ limit of the massive flavour contribution to the EE in flat space, as computed recently in \cite{Jones:2015twa}.     This is not surprising in view of the fact that in the large mass limit we are integrating only over the part of AdS close to the boundary and we have used the leading order embedding (\ref{eqn:perturb large}) which is the same as the flat space embedding (although the coordinate transformation differs from the flat case).   

It is also interesting that the net (CFT+massive) coefficient of $\ln{(R/l)}$ seems to be $0$ by virtue of (\ref{eqn:semi1}).   This agrees with the argument in \cite{maldacenapimentel} that when the bulk geometry is gapped this term should be zero.

\section{Conclusions}  
\label{sec:conclude}
In this work we have used a probe brane holographic model to study the entanglement entropy (EE) of spherical regions in 4-dimensional de Sitter spacetime, for a strongly coupled CFT deformed by massive flavour fields.     We have focussed on the massive part of the flavour contribution to the EE.  There is also the original CFT contribution which we looked at in Sec. \ref{sec:cft} as well as a contribution coming from massless flavours, which we worked out to be a constant multiple of the original CFT contribution.    The massive flavour contribution\footnote{Or rather its finite part, to be precise, as it also has a UV divergent piece (\ref{eqn:j}).} has been computed in two different limits, one where the mass, measured in units of $\sqrt{\lambda}/2\pi$, is small compared to the de Sitter scale, $ml\ll 1$, and one where the mass is large, $m l \gg 1$.    Although most of the final integrals required numerical evaluation, we managed to extract more analytical information than was expected.   The results agree with general expectations from field theory and other holographic arguments:    
\begin{itemize}  
\item
When the size of the sphere is smaller than the de Sitter scale $R/l \ll 1$, the EE as a function of $R/l$ tracks the flat space behaviour predicted using a similar probe brane model \cite{Karch:2014ufa}, at least in the small $ml\ll 1$ limit.   
\item
In the limit of large spheres $R \gg l$ we see a contribution proportional to the number of e-foldings, $\ln{(R/l)}$, which measures long range entanglement.    It is present for the CFT and also receives massive corrections.   We only see it in the regime of small masses $m l\ll 1$, corresponding to an ungapped bulk geometry.   When $ml$ is large and the brane embedding is gapped, we see no $\ln{(R/l)}$ term.   This is in complete agreement with the arguments of \cite{maldacenapimentel} that this term is sensitive to the FRW geometry behind the horizon of the de Sitter slicing of AdS, and is absent when the bulk dual is gapped.       
\item
In the limit $ml\gg 1$ and $mR>1$, we find evidence for the universal contribution logarithmic in the mass, that is, of the form $(mR)^2 \ln{(m\epsilon)}$, which is expected to appear whenever a CFT is deformed by a relevant operator.     It comes with the same coefficient as in flat space, which is consistent because this multiplier is known not to depend on the curvature of the background \cite{Hung:2011ta}.    
\end{itemize}
It is slightly puzzling that we only see this universal contribution in the limit of large $m l \gg 1$ and it appears to be absent for small $m l\ll 1$.    In the bulk, the origin of this difference is the constraint $z<1/m$ in the integrals i.e. it is only present in the gapped embeddings.  
A similar phenomenon occurs in flat space\footnote{In \cite{Karch:2014ufa} it was argued that for $mR<1$ this logarithmic term is in fact present if you expand $\ln{(\epsilon/2R)}$ in (\ref{eqn:flatspace}) as $\ln{(m\epsilon)}-\ln{(2mR)}$, however, this is not clear to us.   In the $mR\gg 1$ regime the presence of this term is unambiguous \cite{Jones:2015twa}.} where this logarithmic term (with the same coefficient as ours) is seen only when $m R>1$ and not when $ mR<1$ \cite{Jones:2015twa}.   In the flat case, the bulk embedding is gapped for any $m R$ and the difference between the two cases is that for $mR<1$ the entangling surface in the back-reacted geometry stays close to the boundary and does not reach the point $z=1/m$ where the branes end, whilst this is not so for $m R>1$ and the constraint must be taken into account.   This is exactly the situation we have here with de Sitter in the $ml\gg 1$ limit:   the log term is absent for $m R<1$, and present for $ m R>1$.   Thus it appears that this limit is reproducing the flat space physics.    This is not surprising as the leading order solution (\ref{eqn:perturb large}) that we have used is the same as the flat space one.     

For small $ml \ll 1$ the de Sitter result for small $R/l$ also agrees with flat space behaviour.    However, at very large $R/l$ one could have large $m R$ and so it is a bit surprising that we do not see the $(mR)^2 \ln{(m\epsilon)}$ term here.      Presumably this is an artefact of the perturbative approach: a power series in $ml$ can never give a $\ln{(ml)}$ term.  One could take the exact supersymmetric embedding in \cite{Karch:2015kfa} and compute the flavour EE to verify if such a term can arise in the ungapped case.   

The EE of superhorizon sized spheres in the ungapped $ml \ll 1$ regime raises another interesting question.  There will come some point in the FRW geometry behind the horizon of the de Sitter slicing of AdS where the $S^3$ caps off and the brane ends, so we should not integrate beyond it.   However, we should also not integrate past the maximal scale factor slice in the FRW geometry, since this is as far as the EE of late time superhorizon sized spheres probes.    Which of these happens first?   We have verified that in the perturbative solution $Y_1$ the $S^3$ caps off at $\hat{\xi}_0>\pi L/2$ (i.e. beyond the maximal scale factor slice) provided $m l < 2/\pi$.   Thus for small $m l$ it is consistent even for very large spheres to integrate all the way up to the maximal scale factor slice when computing the EE.   As we increase $ m l$ there must come a point where the $S^3$ caps off before the maximal scale factor slice is reached (since in the limiting case where we are at the critical mass $m_* l $ the $S^3$ caps off at the horizon).\footnote{In our numerical solution we see this switch occurring at $m l \approx 0.91$.  In other words, when $0< m l < 0.91$, the $S^3$ caps off on the far side of the maximal scale factor slice, whilst for $0.91 < m l < m_* l=1.32$ the $S^3$ caps off between the maximal scale factor slice and the horizon.}   This would then provide a non-trivial bound in the integral which would yield a term logarithmic in the mass, precisely at large $R/l$.  This would explain the puzzle above though it would be good to understand it from the field theory perspective as well.   

The idea that for small $m l$ the EE of superhorizon sized regions at late times fails to probe a wedge of the FRW geometry on the brane worldvolume between the maximal scale factor slice and the point where the $S^3$ caps off is reminiscent of similar situations where geometric field theory probes of a bulk singularity are bounded away from the singularity, see e.g. \cite{Engelhardt:2013tra}.  Examples involving de Sitter field theories were recently discussed in \cite{Kumar:2015jxa,Kumar:2015gln}.  

It would be interesting to study the behaviour of the flavour EE close to the phase transition at the critical $m_* l$.   One could do this with the exact solution in \cite{Karch:2015kfa} but one would have to take into account the contribution of the worldvolume gauge field.   Other possible extensions would be to other probes in de Sitter sliced AdS (e.g. D5 branes modelling flavours on a codimension-1 surface), other shapes of entangling region, and entanglement between two disjoint, separated regions.    The case of a single spherical region is somewhat special in that one can apply the CHM map and make use of the recipe of \cite{Karch:2014ufa} to compute the EE just from the probe embedding.   However, more general methods are known that are applicable to non-spherical regions, for example \cite{Chang:2013mca} that expresses the change in EE as a convolution of the linearised backreaction in the five-dimensional metric and an effective energy momentum tensor, integrated over the original entangling surface (which is simpler than the full backreaction problem).

\acknowledgments 
I would like to thank Prem Kumar for stimulating discussions, Karta Kooner for contributions in the early phase of this project, and Juan Pedraza for helpful correspondence.  
This work was supported in part by the STFC grant award ST/K502376/1.

\appendix

\section{CFT entanglement entropy from RT}
\label{app:RT}
We will show how to reproduce the results in Sec. \ref{sec:cft} by applying the Ryu-Takayanagi (RT) formula \cite{Ryu:2006bv} which expresses the holographic EE in terms of the minimal surface and its covariant generalisation \cite{Hubeny:2007xt} to extremal surfaces.   It is easiest to derive the equation of the minimal surface in the global slicing of AdS, and then transform it to the de Sitter slicing.   For a good account of the minimal surfaces in empty AdS corresponding to spherical entangling regions on the boundary, see \cite{Krtous:2013vha}.  

\subsection*{Global AdS foliation} 
The global foliation of $AdS_{d+1}$ is
\begin{align}
y_{-1}&=L\sec{\chi} \cos{T}\quad && y_0 = L\sec{\chi}\sin{T} \nonumber \\ 
y_i&= L\tan{\chi}\,n_i \sin{\theta}\quad && y_d = L\tan{\chi}\cos{\theta}
\end{align}
with $0\leq \chi \leq \pi/2$, $-\infty<T<\infty$ and the line element
\begin{equation}
ds^2 = L^2\sec^2{\chi}\left(-dT^2+d\chi^2+\sin^2{\chi}\,\left(d\theta^2+\sin^2{\theta}\,d\Omega_{d-2}^2\right)\right)
\end{equation}
The boundary at $\chi=\pi/2$ has geometry $\mathbb{R}\times S^{d-1}$ and is static.   The EE of a spherical region on a constant time slice $T=T_0$ is found by solving for a minimal surface in the $H^d$ spatial section that wraps the $S^{d-2}$ and extends along the $\chi,\theta$ directions 
\begin{equation}
\mathcal{S} = \frac{V_{S^{d-2}}}{4G_N}\int \left(L\sec{\chi}\right)^{d-1}\sin^{d-2}{\chi}\,\sin^{d-2}{\theta}\,\sqrt{d\chi^2+\sin^2{\chi}\,d\theta^2}
\end{equation}
The task is made easier by switching from the $S^{d-1}$ slicing of $H^d$ to a ``cylindrical" slicing where another isometry is manifest \cite{Krtous:2013vha}.    If we let
\begin{equation}
\tanh{\zeta} = \sin{\chi}\cos{\theta},\qquad \cosh{\xi}=\sqrt{\sec^2{\chi}-\tan^2{\chi}\cos^2{\theta}}
\end{equation}
the spatial metric becomes
\begin{equation}
ds_{spatial}^2 = L^2\left(\cosh^2{\xi}\,d\zeta^2+d\xi^2+\sinh^2{\xi}\,d\Omega_{d-2}^2\right)
\end{equation}
and it simply follows that the hyperplanes $\zeta=\zeta_0$ provide a one-parameter set of minimal surfaces with the required symmetry.   In the original coordinates they correspond to
\begin{equation}
T=T_0,\qquad \sin{\chi}\cos{\theta} =\cos{\theta_0}
\end{equation}
where we defined $\cos{\theta_0} \equiv \tanh{\zeta_0}$.

\subsection*{Static de Sitter foliation}
In the metric (\ref{eqn:static slicing metric}) the EE of a spherical region on a constant time slice $t=t_0$ is found by solving for a minimal surface in the spatial metric that wraps the $S^{d-2}$ and extends along the $r,z$ directions.   The area functional to be minimised is
\begin{equation}\label{eqn:smin}
\mathcal{S} = \frac{V_{S^{d-2}}}{4G_N} \int \left(\frac{L}{z}\right)^{d-1} \left(r\sqrt{f(z)}\right)^{d-2} \sqrt{dz^2+\frac{f(z)}{1-r^2/l^2}dr^2}
\end{equation}
where $f(z) = (1-z^2/4l^2)^2$ as before.  The solution for $r(z)$ with the boundary condition $r(0)=R$ is \cite{Fischler:2013fba}
\begin{equation}\label{eqn:exactsolstatic}
r(z) = \sqrt{R^2 - \frac{z^2(1-R^2/l^2)}{f(z)}}
\end{equation}
In fact if we do the coordinate transformations it can be seen that this is the same minimal surface as the one we obtained in global AdS, upon identifying
\begin{equation}
\tan{T_0} = \sqrt{1-R^2/l^2}\sinh{(t_0/l)},\qquad \cos^2{\theta_0} = \frac{\left(1-R^2/l^2\right)\cosh^2{(t_0/l)}}{\left(1-R^2/l^2\right)\cosh^2{(t_0/l)}+R^2/l^2}
\end{equation}
The coordinate $r$ gets smaller as we move deeper into the bulk until it caps off at  
\begin{equation}\label{eqn:zmax}
\frac{z_{max}}{2l} = \frac{1-\sqrt{1-R^2/l^2}}{R/l}
\end{equation}
To find the EE we integrate (\ref{eqn:smin}) between the cut-off at $z=\epsilon$ and $z=z_{max}$.   For the CFT in $dS_4$ this gives the EE 
\begin{equation}\label{eqn:cft4}
\mathcal{S}_{CFT_4} = \frac{L^3 V_{S^2}}{16 G_N}\left(\frac{2 R^2}{\epsilon^2}-2\ln{(2 R/\epsilon)}+\frac{R^2}{l^2}-1\right)
\end{equation}
in agreement with (\ref{eqn:eecft4}).   One could argue that we should impose the boundary condition $r=R$ on the regulator surface $z=\epsilon$.   It turns out this is equivalent to sending $R^2 \to R^2 - \left(1-R^2/l^2\right)\epsilon^2$ and therefore its net effect is to reverse the sign of the finite piece in (\ref{eqn:cft4}).   This explains the discrepancy between (\ref{eqn:eecft4}) and \cite{Fischler:2013fba}.    It illustrates the fact that the finite piece is scheme-dependent.

\subsection*{Global de Sitter foliation}
Looking at the metric (\ref{eqn:global slicing metric}) the boundary is not static, so the problem involves extremal surfaces that extend in the time direction and requires one to use the covariant EE of \cite{Hubeny:2007xt}.  The EE of a spherical region on a constant time slice $\tau=\tau_0$ is found by solving for an extremal surface in the metric that wraps the $S^{d-2}$ and extends along the $\tau,\theta,z$ directions.   The area functional to be extremized is 
\begin{equation}
\mathcal{S} = \frac{V_{S^{d-2}}}{4G_N}\int \left(\frac{L}{z}\right)^{d-1}\left( l \cosh{(\tau/l)}\,\sin{\theta}\sqrt{f(z)}\right)^{d-2} \sqrt{dz^2 - f(z)d\tau^2+f(z)\,l^2\cosh^2{(\tau/l)}d\theta^2}
\end{equation}
Instead of solving this we can take the minimal surface we found in global AdS and transform it to the global dS slicing.    In embedding coordinates the surface takes the form
\begin{equation}
\frac{y_0}{y_{-1}} = \tan{T_0},\qquad \frac{y_d}{\sqrt{y_{-1}^2+y_0^2}} = \cos{\theta_0} 
\end{equation} 
which in the global dS slicing means
\begin{equation}
\frac{1-z^2/4l^2}{1+z^2/4l^2}\sinh{(\tau/l)} = \sinh{(\tau_0/l)},\qquad \coth{(\tau/l)}\cos{\theta}=\cos{\theta_0}\,\coth{(\tau_0/l)} 
\end{equation}
where $\tan{T_0} = \sinh{(\tau_0/l)}$ to ensure that $\tau=\tau_0,\,\theta=\theta_0$ at the boundary $z=0$.    Again we can evaluate the area integral for this solution to find the EE.   It reduces to the same integral as we had to do in static de Sitter with the replacement $R = l \sin{\theta_0} \cosh{(\tau_0/l)}$.  The surface ends in the bulk at $z_{max}$ given by (\ref{eqn:zmax}).   When $R<l$ it ends before reaching the bulk horizon at $z=2l$ and we again get (\ref{eqn:cft4}) for the CFT in $dS_4$.   When $R>l$ the surface extends all the way through the horizon in the bulk,
capping off at a complex $z_{max}$.    As discussed in Sec. \ref{subsec:horizon} this corresponds to the extremal surface probing the FRW geometry behind the horizon.     Computing the integral with the formally complex upper limit $z_{max}$ we again get (\ref{eqn:cft4}), in agreement with (\ref{eqn:eecft4}).

\section{Metric determinant and variation}
\label{app:metric}
Whilst it is possible to write the metric explicitly in terms of the coordinates $(\hat{\sigma},\rho,u)$, it is non-diagonal and its determinant and variation do not easily yield to simplification, so it is better to use other ways.    To evaluate the determinant of the induced metric  (\ref{eqn:induced metric hyperbolic}), we write it in the original coordinates (\ref{eqn:induced metric}), factor out the $\sqrt{g_{AdS_5}}$ part and transform that part to the hyperbolic slicing (effectively multiplying by the Jacobian):
\begin{align}
\sqrt{\gamma} &= \sqrt{g_{AdS_5}} \sqrt{1+z^2 \psi'^2}\, L^3 \sin^3{\psi} \sqrt{g_{S^3}}  \nonumber \\
& \to \rho^3 \sinh^2{u}\,\sqrt{g_{S^2}}\, \sqrt{1+z^2 \psi'^2}\, L^3 \sin^3{\psi} \sqrt{g_{S^3}}
\end{align} 
where $\psi_z=\psi'(z)$.   To evaluate the variation of the determinant we write the metric for the replica geometries (\ref{eqn:replica metric}) as the sum of the $n=1$ metric and the first variation
\begin{align}
\gamma_n &= g_{AdS_5}^{hyp,n} +  L^2 \psi'^2(z(\hat{\sigma},\rho,u)) dz(\hat{\sigma},\rho,u)^2 + L^2 \sin^2{\psi(z(\hat{\sigma},\rho,u))}\,d\Omega_3^2 \nonumber \\
g_{AdS_5}^{hyp,n} & = g_{AdS_5}^{hyp} + \frac{2L^2}{3\rho^2}\left(d\hat{\sigma}^2 - \frac{d\rho^2}{\left(\frac{\rho^2}{L^2}-1\right)^2}\right)(n-1)+\mathcal{O}\left((n-1)^2\right)
\end{align} 
Now using $\delta\sqrt{g} = \frac{1}{2}\sqrt{g}\,g^{m\nu}\delta g_{m\nu}$ it follows that 
\begin{equation}
\partial_n \sqrt{\gamma_n}|_{n=1} = \frac{1}{2}\sqrt{\gamma} \gamma^{m\nu} \delta\gamma_{m\nu} = \frac{1}{2}\sqrt{\gamma}\left(\frac{2L^2}{3\rho^2}\right)\left(\gamma^{\hat{\sigma}\hat{\sigma}} - \frac{\gamma^{\rho\rho}}{\left(\frac{\rho^2}{L^2}-1\right)^2}\right)
\end{equation}
The two components of the $n=1$ inverse metric are
\begin{align}
\gamma^{\hat{\sigma}\hat{\sigma}} &=  \left(-\gamma_{\rho u}^2+\gamma_{\rho\rho}\gamma_{uu}\right) \left(\rho^2 \sinh^2{u}\,\sqrt{g_{S^2}}(L\sin{\psi})^3\,\sqrt{g_{S^3}}\right)^2 \gamma^{-1} \nonumber \\
\gamma^{\rho\rho} &= \left(-\gamma_{\hat{\sigma} u}^2+\gamma_{\hat{\sigma}\hat{\sigma}}\gamma_{uu}\right) \left(\rho^2 \sinh^2{u}\,\sqrt{g_{S^2}}(L\sin{\psi})^3\,\sqrt{g_{S^3}}\right)^2 \gamma^{-1}
\end{align}
where 
\begin{align}
\gamma_{\rho u} &=  L^2 \psi'^2 \partial_\rho z\, \partial_u z,\quad &&\gamma_{\rho\rho} =\left(\frac{\rho^2}{L^2}-1\right)^{-1} + L^2 \psi'^2 (\partial_\rho z)^2,\,\, &&& \gamma_{uu}= \rho^2  + L^2 \psi'^2 (\partial_u z)^2 \nonumber \\
\gamma_{\hat{\sigma} u} &=  L^2 \psi'^2 \partial_{\hat{\sigma}} z\, \partial_u z,\quad && \gamma_{\hat{\sigma}\hat{\sigma}} = \frac{\rho^2}{L^2}-1 +  L^2 \psi'^2 (\partial_{\hat{\sigma}} z)^2
\end{align}
The variation simplifies to 
\begin{equation}
\partial_n \sqrt{\gamma_n}|_{n=1} = \frac{\sqrt{\gamma}\, L^2 \psi_z^2}{1+z^2\psi_z^2}\frac{\left(\frac{\rho^2}{L^2}-1\right)^2(\partial_\rho z)^2 - (\partial_{\hat{\sigma}} z)^2}{\frac{3\rho^2}{L^2}\left(\frac{\rho^2}{L^2}-1\right)^2}
\end{equation}

\end{document}